\documentclass{article}
\usepackage{graphicx}
\usepackage{fullpage}
\usepackage{subcaption}
\usepackage{todonotes}
\usepackage{amsmath}
\usepackage{amsfonts}
\usepackage{natbib}
\usepackage[symbol]{footmisc}

\makeatletter
\def\ps@pprintTitle{%
  \let\@oddhead\@empty
  \let\@evenhead\@empty
  \def\@oddfoot{\footnotesize\itshape Preprint}%
  \let\@evenfoot\@oddfoot
}
\makeatother

\title{Bridging Cognitive Maps: a Hierarchical Active Inference Model of Spatial Alternation Tasks and the Hippocampal-Prefrontal Circuit}

\author{Toon Van de Maele  \\
	IDLab, Department of Information Technology, Ghent University - imec, Ghent, Belgium  \\
	\and 
	Bart Dhoedt \\
	IDLab, Department of Information Technology, Ghent University - imec, Ghent, Belgium  \\
 \and
Tim Verbelen$^*$  \\
	VERSES Research Lab, Los Angeles, USA  \\
\and
Giovanni Pezzulo$^*,\dag$  \\
	Institute of Cognitive Sciences and Technologies, National Research Council, Rome, Italy  \\
	}

\footnotetext{These authors contributed equally to this work}
\footnotetext{Corresponding author: giovanni.pezzulo@istc.cnr.it}

\footnotetext{\\\normalsize{\textit{Preprint}}}

\begin{document}
\maketitle

\begin{abstract}

Cognitive problem-solving benefits from cognitive maps aiding navigation and planning. Previous studies revealed that cognitive maps for physical space navigation involve hippocampal (HC) allocentric codes, while cognitive maps for abstract task space engage medial prefrontal cortex (mPFC) task-specific codes. Solving challenging cognitive tasks requires integrating these two types of maps. This is exemplified by spatial alternation tasks in multi-corridor settings, where animals like rodents are rewarded upon executing an alternation pattern in maze corridors. Existing studies demonstrated the HC – mPFC circuit's engagement in spatial alternation tasks and that its disruption impairs task performance. Yet, a comprehensive theory explaining how this circuit integrates task-related and spatial information is lacking. We advance a novel hierarchical active inference model clarifying how the HC – mPFC circuit enables the resolution of spatial alternation tasks, by merging physical and task-space cognitive maps. Through a series of simulations, we demonstrate that the model's dual layers acquire effective cognitive maps for navigation within physical (HC map) and task (mPFC map) spaces, using a biologically-inspired approach: a clone-structured cognitive graph. The model solves spatial alternation tasks through reciprocal interactions between the two layers. Importantly, disrupting inter-layer communication impairs difficult decisions, consistent with empirical findings. The same model showcases the ability to switch between multiple alternation rules. However, inhibiting message transmission between the two layers results in perseverative behavior, consistent with empirical findings. In summary, our model provides a mechanistic account of how the HC – mPFC circuit supports spatial alternation tasks and how its disruption impairs task performance.\\


\end{abstract}






\newpage
\section{Introduction}

To solve cognitive problems, such as finding the shortest route to a goal destination in a busy city, we can use so-called \emph{cognitive maps} of the situation that affords flexible planning. The concept of the cognitive map has been initially popularized by Tolman, especially in the context of spatial navigation \citep{tolman1948cognitive}. An extensive body of literature has investigated the neuronal underpinnings of cognitive maps for spatial navigation in humans, rodents, and other animals. This research has established the crucial importance of structures of the medial temporal lobe and especially of the hippocampal formation in the creation of codes for allocentric space, such as place cells (in the hippocampus) \citep{okeefe_precis_1979,OKeefe1971} and grid cells (in the entorhinal cortex) \citep{hafting2005microstructure}. An emerging perspective to understand these findings is that cognitive mapping in the hippocampal formation might be described at the computational level as a sequence learning problem and that place cells, grid cells and other specialized codes might emerge from such sequence learning \citep{whittington_tolman-eichenbaum_2020,whittington_how_2022,george_clone-structured_2021,raju2022space,stachenfeld2017hippocampus,stoianov_hippocampal_2022,chen2022predictive,levenstein2024sequential,recanatesi2021predictive}. Furthermore, it has been suggested that the same hippocampal codes that support spatial navigation might also support navigation in ``cognitive'' domains \citep{bellmund2018navigating,epstein2017cognitive,buzsaki2013memory}.

In parallel, the concept of a cognitive map can apply to the abstract state spaces that describe the sequential stages of a task at hand (for example, buying a gift in one's preferred shop and then bringing it to a friend's house) as opposed to the more fine-grained spatial codes found in the hippocampal formation. Recent studies reported that prefrontal structures, such as the orbitofrontal cortex, might host such coarse task codes, which might permit representing (for example) the current and the next stages of the task \citep{niv2019learning,schuck_human_2016} as well as future navigational goals \citep{basu2021orbitofrontal}, as opposed to a physical location.  

Crucially, during goal-directed spatial navigation (and other cognitive problems), it is often necessary to combine cognitive maps of task space (to find a sequence of coarse-grained actions to progress in task space, such as ensuring the correct sequence of destinations to buy and deliver a gift) and physical space (to find a path that reaches the next intended destination) \citep{ito2015prefrontal,pezzulo2014principles,verschure2014and}. However, we still lack a comprehensive theory that explains the ways cognitive maps of physical and task space interact when executing cognitive tasks.

\begin{figure}
        \begin{subfigure}{\textwidth}
            \centering
            \includegraphics[width=0.80\textwidth]{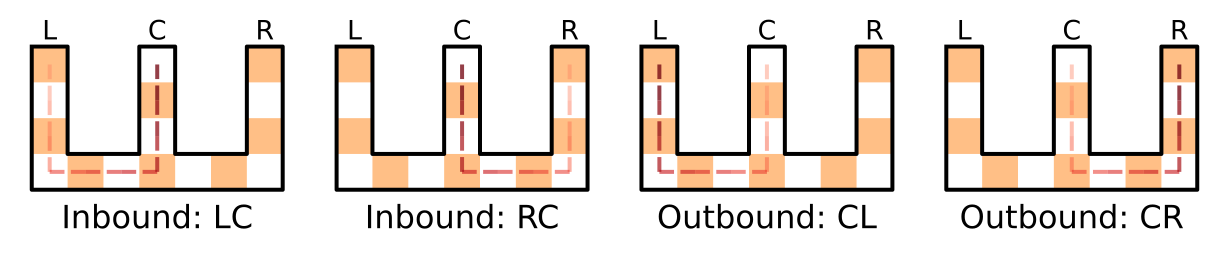}
            \subcaption[]{}
        \end{subfigure}
        \caption{\textbf{Spatial alternation tasks (LCRC).} This figure exemplifies the W-maze used in ~\citep{jadhav_awake_2012} to study spatial alternation and the role of the hippocampal - prefrontal dynamics. The W-maze is characterized by three separate corridors. The animal can acquire the reward at the end of each corridor, however, to receive the reward it must do so in the correct order (e.g. \textbf{L}eft, \textbf{C}enter, \textbf{R}ight, \textbf{C}enter, \textbf{L}eft, etc). The spatial alternation task implies that when the animal is in the center position, it has to make a (difficult) \emph{outbound} decision: it has to go either to the left or the right, depending on where it comes from. This outbound decision, therefore, requires a spatial memory component. In contrast, when the animal is in one of the side corridors (left or right), the only correct action is to move to the center. This \emph{inbound} decision is considered simpler since it does not require a memory component. Trajectory order is indicated by the shade where lighter is earlier.}
        \label{fig:spatial_alternation}
    \end{figure}

A useful starting point to understand the interaction of task-related and spatial codes is rodent memory-guided, spatial alternation tasks. A prominent example is the spatial alternation task in the W-maze studied in~\citep{jadhav_awake_2012}. As visualized in Figure~\ref{fig:spatial_alternation}, the W-maze comprises three corridors. To collect rewards, the animal has to visit the corridors in the correct order, according to an ``alternation rule'' (e.g. left, center, right, center, left, etc) that is initially unknown and has to be learned by trial and error. 

A series of experiments showed that solving the spatial alternation task engages coordinated patterns of neural activity in the hippocampus (HC), which putatively encodes the spatial aspects of the task, and the prefrontal cortex (mPFC), which putatively learns the alternation rule and uses it to guide behavior~\citep{benchenane_oscillations_2011,shin_multiple_2016,colgin2011oscillations,siapas2005prefrontal,khodagholy2017learning}. For example, the HC - mPFC interaction is selectively enhanced during epochs requiring spatial working memory~\citep{jones_theta_2005}. Furthermore, the coordination of neural activity in both structures spans multiple timescales, from theta sequences during navigation to reactivation (replay) activity during inter-trial periods prior to navigation -- and is considered crucial for the information exchange between the two areas during the task~\citep{tang_multiple_2021}. 

Notably, disrupting awake hippocampal reactivations (sharp wave ripples, SWR) at decision points impaired the animal's performance, but only for the aspects of the alternation task that require spatial working memory~\citep{jadhav_awake_2012}, see Figure \ref{fig:spatial_alternation}. Specifically, disrupting awake SWRs impaired \emph{outbound} decisions (from the central corridor to the left or right corridor), which require memory for immediate past outer arm location. Rather, the same SWR disruption did not impair \emph{inbound} decisions (from the left or right corridor to the central corridor) that do not require memory of the past corridor; nor did it impair self-localization~\citep{jadhav_awake_2012}. The study concludes that impairing SWRs prevents the hippocampus from providing information about past locations (and/or future options) to the prefrontal cortex, which is responsible to learn the alternation rule and to use it to guide behavior. In other words, the SWR disruption impairs communication between the hippocampus and the prefrontal cortex, hence preventing the latter from correctly inferring the current stage in task space. 

The importance of hippocampal-prefrontal communication for memory guided decisions is confirmed by another study in which animals had to learn and subsequently switch between three spatial alternation rules in the same three-arm maze ~\citep{den_bakker_sharp-wave_2022}. This study showed that disrupting neural activity in the mPFC directly following hippocampal sharp-wave ripples (but not after a random delay) impairs the animal's ability to switch between learned rules.

Here, we advance a novel computational theory that casts the interactions between the hippocampus (HC) and the medial prefrontal cortex (mPFC) during spatial alternation tasks, in terms of hierarchical active inference \citep{friston_free-energy_2010,friston_active_2017,bogacz_tutorial_2017,buckley_free_2017,parr_active_2022,smith_step-by-step_2022,isomura_experimental_2023}. Our theory has two main tenets. First, the HC and the mPFC learn cognitive maps for navigation in physical and task space, respectively, using the same statistical computations modeled as Partially Observable Markov Decision Processes, POMDP~\citep{kaelbling1998planning}), but based on different inputs (see Section~\ref{sect:method}). Second, the neural circuit formed by the HC and the mPFC realizes a hierarchical active inference architecture, which can solve spatial alternation tasks. Hierarchical active inference rests on reciprocal, bottom-up, and top-down message passing. In the bottom-up pathway, the lower hierarchical level (encoding the HC map) infers the current location based on sensory information and communicates it to the higher level (encoding the mPFC map). In turn, the higher level infers the next goal location (in the mPFC map) and sets it as a goal for spatial navigation (in the HC map), in a top-down manner.  Impairments of the message passing prevent the architecture from correctly solving spatial alternation tasks. The separation of timescales automatically emerges from the structure of hierarchical active inference. The lower level passes the bottom-up message when a free-energy threshold is met, indicating it is confident it has reached its goal.

We exemplify the new theory by presenting two simulations of rodent spatial alternation tasks, with intact and impaired HC- mPFC interactions ~\citep{jadhav_awake_2012,den_bakker_sharp-wave_2022}. Our first simulation, presented in Section~\ref{sect:sim1}, brings three main conclusions. First, the model is able to learn the spatial structure of the maze (HC map) and spatial alternation rules (mPFC) by navigating in the environment. Second, when the HC - mPFC circuit is configured as a hierarchical active inference system, it effectively solves spatial alternation tasks, through bottom-up and top-down message passing. Third, disrupting the interaction from the HC to the mPFC breaks the spatial memory and impairs the animal's ability to make outbound but not inbound decisions, as observed empirically~\citep{jadhav_awake_2012}. Our second simulation, presented in Section~\ref{sect:sim2}, shows two additional features of the model. First, in a task in which three alternative spatial alternation rules are in play, the HC - mPFC circuit permits inferring the current rule. Second, selectively inhibiting the mPFC impairs this ability and provokes distinct behavioral patterns, as observed empirically in the rodent study reported in~\citep{den_bakker_sharp-wave_2022}.

We provide the code for the simulations at https://github.com/toonvdm/bridging-cognitive-maps.
    
\section{Results}
 
    \subsection{The W-maze setup and the hierarchical active inference (HAI) agent}

    In this work, we consider the spatial alternation task in the W-maze of~\citep{jadhav_awake_2012}, see Figure~\ref{fig:spatial_alternation}. In this task, an animal (or an artificial agent) can acquire rewards at the end of each corridor, providing that the corridors are visited in the correct order (e.g. left, center, right, center, left).

    In this section, we present a hierarchical active inference (HAI) agent \citep{parr_active_2022} for solving the spatial alternation task. Central to this framework is the fact that agents are endowed with a generative model, describing how observed outcomes are generated from a hidden, unobserved state, which is influenced by the agent's actions. The agent infers these hidden states by minimizing the free energy functional with respect to its belief over the state and acts to minimize its expected free energy over time~\citep{parr_active_2022}. For more details on these mechanisms, the reader is referred to Section~\ref{sect:hai}.
    
\begin{figure}[h!]
        \centering
        \begin{subfigure}{\textwidth}
            \centering
            \includegraphics[width=0.68\textwidth]{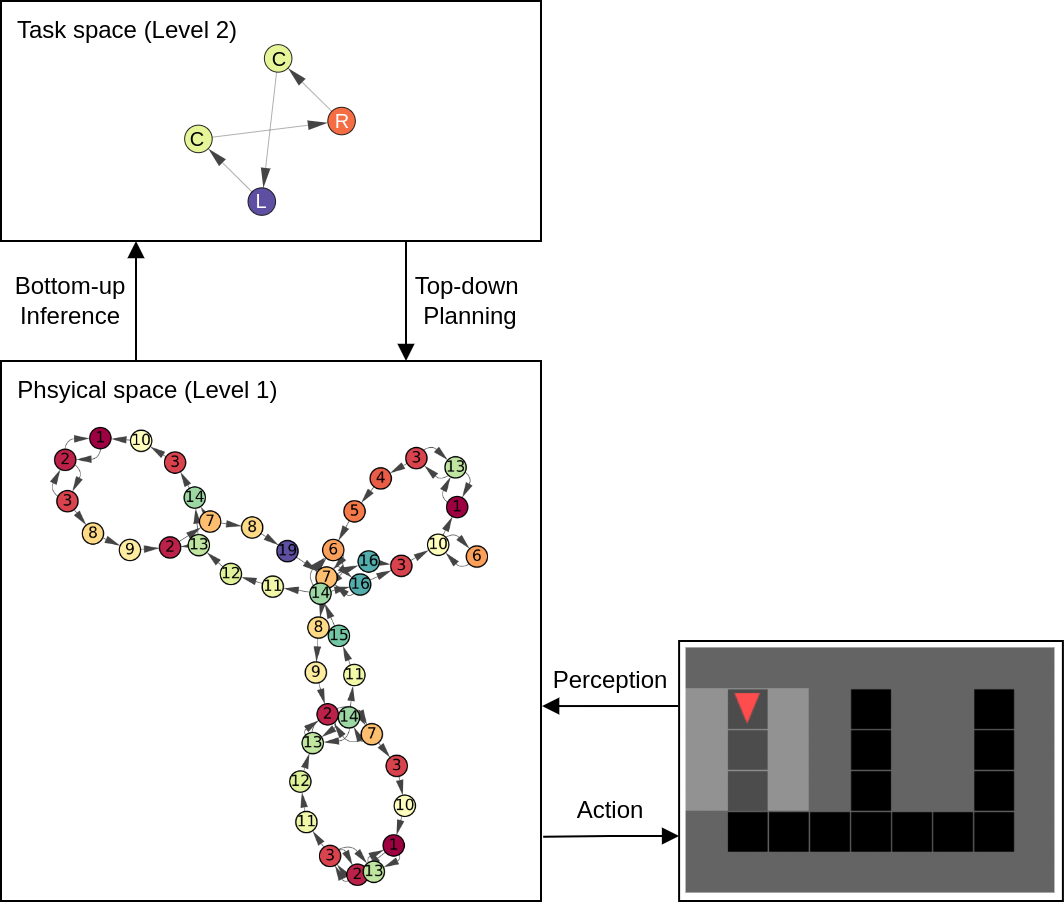}
            \subcaption[]{\label{fig:dynamics}}
            \includegraphics[width=0.60\textwidth]{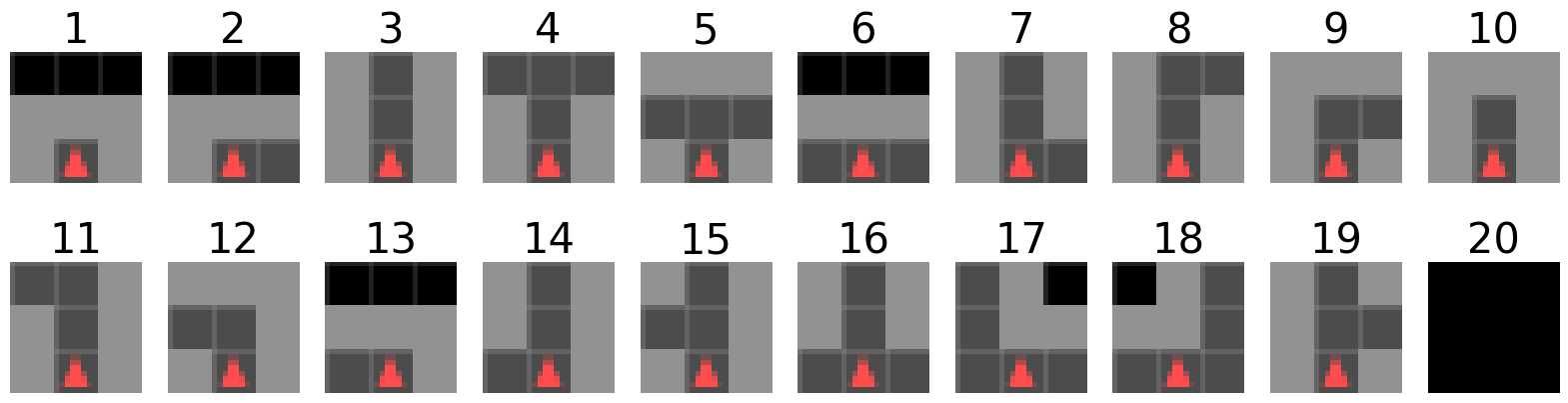}
            \subcaption[]{\label{fig:observations}}
            
        \end{subfigure}
        \caption{\textbf{Illustration of the hierarchical active inference (HAI) agent and the W-maze setup} (a) The hierarchical generative model of the HAI agent. The figure shows that the two layers of the model include cognitive maps of physical space (level 1) and task space (level 2). These cognitive maps essentially represent transitions between learned locations and stages of the task, respectively. Both maps are learned using a probabilistic sequence learning algorithm: the clone structured cognitive graph (CSCG~\citep{george_clone-structured_2021}). Within each block, a subset of states from the CSCG is visualized, namely, the active states when pursuing the alternation rule (LCRC). For a visualization of the full set of states from the CSCG, we refer to Figure~\ref{fig:learned_graphs}. The dynamics between the physical (level 1) and task space (level 2) CSCG in the hierarchical model are regulated by bottom-up message passing (that supports inference) and top-down message passing (that supports planning). Bottom-up inference messages from the physical space model are passed to the task space model, while top-down actions drive planning from the task-space model to the physical space model. An essential aspect of these interactions is that the two levels operate on a different temporal scale. Transitions at the physical level, through the agent's movement, occur at every time step. Rather, the task level only transitions when the intermediate goal (of reaching a certain position in the maze) is achieved. This abstraction allows for hierarchical planning. Finally, the interaction with the environment using an action-perception loop with the world, the W-maze, is shown on the right. (b) The 20 distinct observations the agent can encounter in the W-maze created in the Minigrid environment~\citep{minigrid}. 
        }
        \label{fig:simulation1}
    \end{figure}

    In our simulations, the HAI agent is endowed with a hierarchical generative model of two layers, which is schematically illustrated in Figure \ref{fig:dynamics} (see Section~\ref{sect:hai} for a formal introduction). In this generative model, the lowest layer takes up the functional role of the hippocampus, namely encoding the spatial structure of the maze~\citep{okeefe_precis_1979}. In contrast, the highest layer takes the functional role of the prefrontal cortex, namely encoding the structure of task space~\citep{schuck_human_2016}. The agent learns both maps using the same sequence learning algorithm: the clone structured cognitive graphs (CSCG~\citep{george_clone-structured_2021}), see Section~\ref{sect:method} for details. The latent states learned by the CSCG at the two levels are schematically illustrated within the two boxes of Figure \ref{fig:dynamics}.     

    Crucially, the HAI agent performs inference (of where it is in physical space, i.e., its pose, and in task space) and planning (of the next goal in physical and task space) through message-passing between the two hierarchical levels. As shown in Figure~\ref{fig:dynamics}, level 1 observes the structural aspects of the environment through its bottom-up observations, i.e. the three-by-three grid around the agent as shown in Figure~\ref{fig:observations} and selects local movements in physical space such as ``turn left'', ``turn right'' or ``move forward'' to reach any goal location (which is fed by the highest level, see below). 
    
    Once level 1 has navigated towards its local goal, it passes a bottom-up message containing the current state of the agent to level 2. In turn, level 2 processes this message and infers the next stage of the task to achieve a reward (as it is endowed with a prior preference to achieve reward at each step, see Section~\ref{sect:hai}). Actions at level 2 are abstract and are matched to the bottom-up received observable states from level 1. Since level 1 states encode the agent's pose, these actions essentially mean ``move to target location''. We specifically encode only actions corresponding to target locations at the end of a corridor (i.e. observation 1 in Figure~\ref{fig:observations}). The selected action is sent to level 1 in a top-down manner and corresponds to the goal location for level 1. Once level 1 reaches the goal location, this process repeats. 
    
    Finally, the right part of Figure~\ref{fig:dynamics} illustrates the fact that the lowest level of the model is responsible for the action-perception loop with the environment -- here, an implementation of the W-maze in the Minigrid~\citep{minigrid} simulator. Observations in the Minigrid~\citep{minigrid} are acquired as a square around the agent. In this work, we chose a range of three, yielding three-by-three observations which are then mapped to a unique one-hot index. For each step, the agent also observes a binary value indicating the reward presence. The actions the agent can perform are ``turn left'', ``turn right'' and ``move forward'', which are also represented as a one-hot index. See Section~\ref{sect:hai} for more details on the model implementation


    
    \subsection{Experiment 1: Solving the spatial alternation task}
    \label{sect:sim1}

    
    We replicate the task and experiment from a study on rodents~\citep{jadhav_awake_2012} where the animal is taught this specific alternation task. In this study, disruption of the hippocampal sharp-wave ripple (SWR) reduces the performance on outbound trajectories, but not on inbound trajectories. 

    We first allow the HAI agent to learn the two cognitive maps for solving the spatial alternation task, using two CSCGs (see~\citep{george_clone-structured_2021} and Section~\ref{sect:method} for details on the learning procedure). The learned maps for the W-maze are shown within the two boxes of Figure \ref{fig:dynamics}. For ease of visualization, within each box, only a subset of states from the CSCG is shown, namely, the states that are active when correctly pursuing the alternation rule (the complete learned maps are shown in Figure \ref{fig:learned_graphs}). We denote the sequence of this alternation rule as LCRC (for Left, Center, Right, Center). For a visualization of the full set of states from the CSCG, please see Figure~\ref{fig:learned_graphs}.
    
    The hippocampal map on the lowest level organizes the 20 observations that the agent encounters during navigation (illustrated in Figure~\ref{fig:observations}) into a coherent graph, which nicely reflects the 3 corridors of the W-maze. In this map, the circles reflect the states learned by the CSCG, while the transitions reflect the agent's spatial actions (e.g., ``turn left'', ``turn right'', or ``move forward''). The nodes are color-coded (and numbered) according to the observations that they encode. Note that while the agent encounters the same observations multiple times during navigation (i.e., the observations are aliased), the CSCG correctly reflects the fact that they are part of different behavioral sequences. For example, observation 1 (that corresponds to the location in which a reward can be collected) appears three times, at the apex of each of the latent sequences that represent each of the three corridors. 
     
    The prefrontal map on the highest level corresponds to a much smaller graph that encodes the task-specific sequence of corridors that the agent has to visit to secure rewards. In this graph, the nodes correspond to (color-coded) corridors in which rewards have been collected and the edges correspond to higher-order actions (``go to corridor''). Note that there are two nodes of the same color: both correspond to the central corridor but in different phases in task space (i.e., after the left and the right corridors, respectively). This indicates that the agent has successfully learned a low-dimensional representation of (or a finite state machine for) the task rule LCRC. 

     \begin{figure}
        \centering
         \begin{subfigure}{0.48\textwidth}
             \centering
             \includegraphics[width=0.95\textwidth]{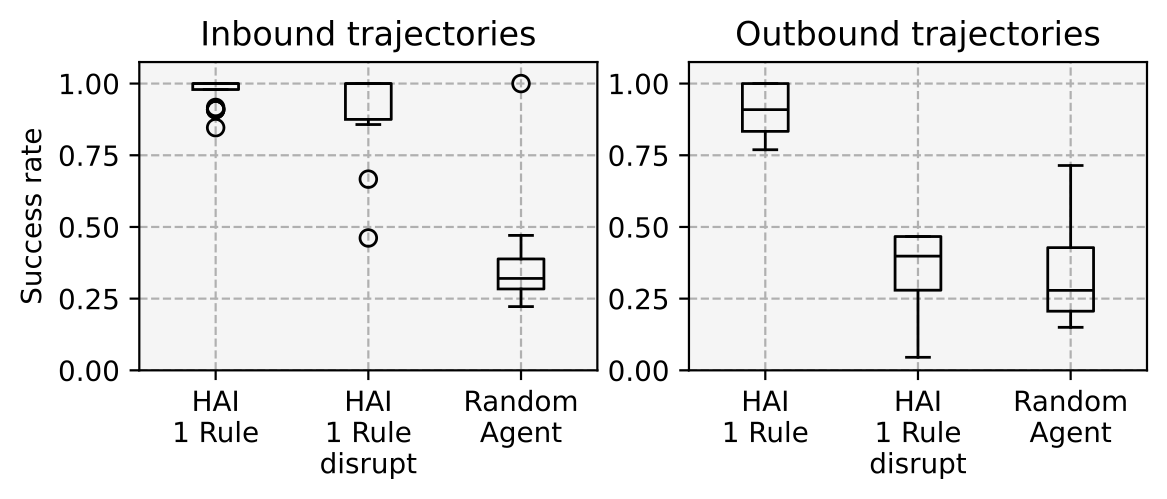}
                \subcaption[]{\label{fig:boxplots}}
        \end{subfigure}
        \begin{subfigure}{.35\textwidth}
            \centering
            \includegraphics[width=0.90\textwidth]{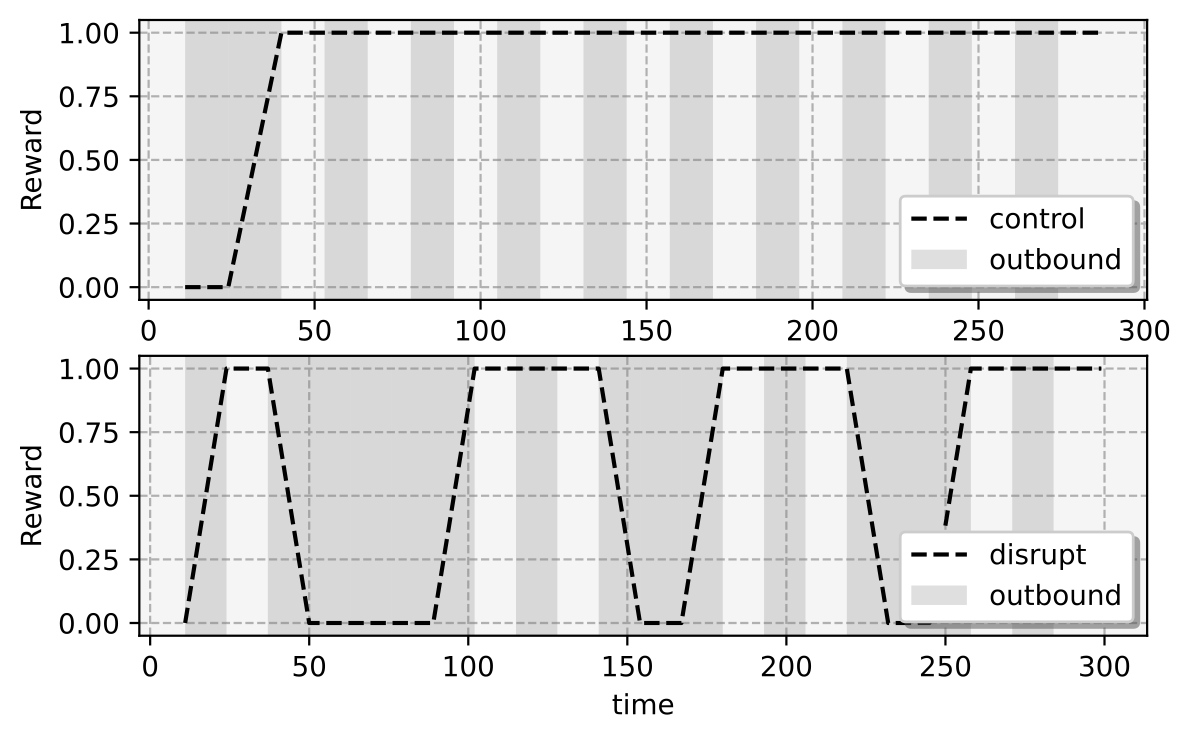}
            \subcaption[]{\label{fig:performance_exp1}}
        \end{subfigure}
        \begin{subfigure}{0.48\textwidth}
            \centering
            \includegraphics[width=0.95\textwidth]{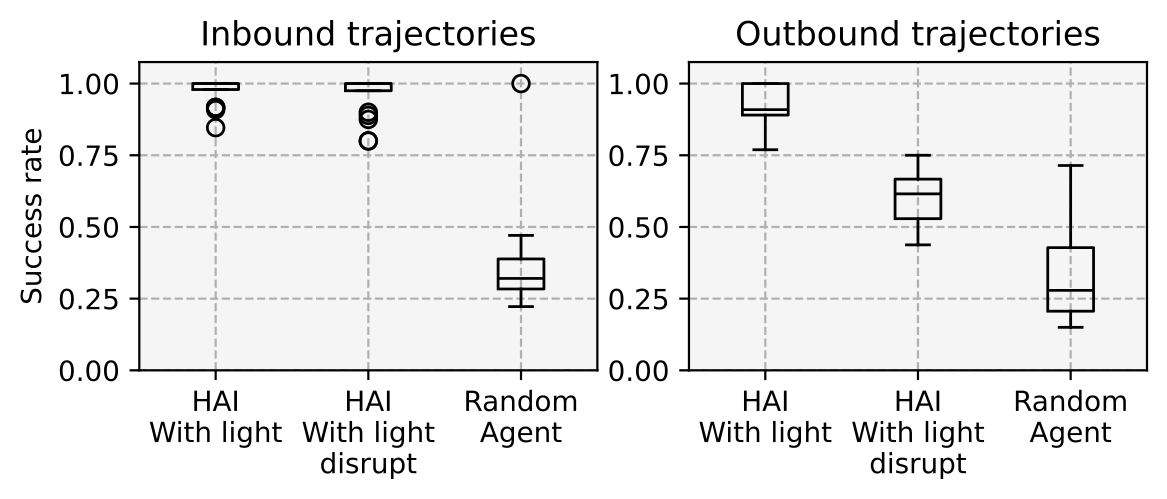}
            \subcaption[]{\label{fig:withlight}}
        \end{subfigure}
        \hfill
        \caption{\textbf{Experiment 1: Solving the spatial alternation task with intact and disrupted spatial memory.} (a) The success rate in the spatial alternation task for the HAI agent without disruption, the HAI agent with disruption (HAI disrupted), and the Random agent, separated for both in- and outbound trajectories. The success rate is computed over 20 trials per agent, with 300 steps per trial, only recording the presence of reward at the end of the selected corridors. (b) The reward over time for the HAI agent and a disrupted HAI agent. The shaded areas in the plot indicate outbound trajectories. (c) The success rate for the HAI agent with (HAI disrupted) and without (HAI) disruption in the alternative light environment. The success rate is computed over 20 trials per agent, with 300 steps per trial, only recording the presence of reward at the end of the selected corridors. The agent is disrupted by disabling the connection from the transition model of the POMDP, separated for in- and outbound trajectories. We observe that the performance on inbound trajectories does not change, while for outbound trajectories this drops to random chance.}
        \label{fig:simulation2}
    \end{figure}

    We then investigate whether the HAI agent equipped with the learned cognitive maps for physical and task space is able to solve spatial alternation tasks. For this, we endow the HAI agent with a prior preference for maximizing the reward and place it in a random pose in the W-maze. We then let the agent run for 300 steps, and record whether it selected the correct or the incorrect corridor to find the reward. We repeat this experiment for 20 trials, to collect statistics. Figure~\ref{fig:boxplots} shows that the HAI agent has near-perfect performance during the easiest, inbound decisions and very high (above 80\%) performance during the harder, outbound decisions, closely approximating the empirical results reported in \cite{jadhav_awake_2012}. Furthermore, as expected, it significantly outperforms a Random agent that selects the next corridor randomly (t-test p=0). Figure~\ref{fig:performance_exp1} shows that after a few time steps, once the HAI agent has inferred where it is (in both physical and task space), it consistently follows the rule and is able to collect rewards at every step.
    
    We performed two control experiments, which reveal that the HAI model is robust to a wide choice of parameters (Appendix ~\ref{sec:model_capacity}) and to greater levels of uncertainty (Appendix ~\ref{sec:noisy_scelario}). Our control experiments also illustrate that in scenarios characterized by greater uncertainty, the expected free energy functional used by the HAI model to select policies is more advantageous compared to the sole objective of reward maximization that is common in economic and reinforcement learning settings. The expected free energy functional effectively balances two components: a pragmatic imperative to maximize reward (e.g., to reach the next subgoal to secure reward) and an epistemic imperative to gain information (e.g., to reduce uncertainty about the current pose, by going to places where unambiguous observations could be found, e.g., the T-junction), see Section~\ref{sect:hai}. By comparing HAI agents with and without the epistemic imperative, we found that the pragmatic imperative to secure rewards is sufficient to address a W-maze with low uncertainty, but adding the epistemic imperative to reduce uncertainty about one's hidden state (here, the pose) significantly increases performance in a W-maze with greater uncertainty (Appendix ~\ref{sec:noisy_scelario}). This is because an agent endowed with epistemic uncertainty can explicitly plan to self-localize (e.g., by visiting unambiguous locations, such as the bottom of corridors), which is especially advantageous with greater uncertainty~\citep{parr_active_2022,schwartenbeck_computational_2019}.

    \subsubsection{Disrupting spatial memory impairs outbound decisions}
    \label{sect:sim2}

    Having established that the HAI agent correctly solves the spatial alternation task, we next ask whether an impairment of spatial memory disrupts its performance in outbound decisions, as shown experimentally by Jadhav et al.~\citep{jadhav_awake_2012}. In this study, the experimenters disrupted hippocampal SWR at decision points and observed that the disruption prevented access to spatial memory -- rendering the rodents unable to make correct outbound decisions -- but left hippocampal spatial codes intact. The experimenters,  therefore, concluded that the disruption was caused by a failure of communication or updating at the level of the PFC. 

    In analogy with this procedure, we realized a variant of the HAI agent (HAI disrupt) in which we implement the SWR disruption by preventing the belief updates (through the transition model) at the second level of the hierarchical model (Figure~\ref{fig:generativemodel} as explained in Section~\ref{sect:hai}). By doing this, we remove the agent's spatial memory about task space. However, the agent is still able to receive the bottom-up message from level 1 and thus knows where it currently is in physical space -- in keeping with the finding that spatial codes are intact after SWR disruptions~\citep{jadhav_awake_2012}. 

When we compare the HAI agent with and without disruption in the spatial alternation task, we observe the same behavior as the experimental study: the HAI disrupted agent correctly addresses inbound decisions but makes outbound decisions at a chance level, see Figure~\ref{fig:boxplots}. The simulation shows that the disruption only impairs the outbound trajectories (please see Appendix~\ref{app:gamma} for additional simulations exploring how the choice of the gamma parameter influences the results). A more detailed example of this pattern of results can be appreciated in Figure~\ref{fig:performance_exp1}, which shows that the HAI disrupted agent tends to miss rewards, but only during outbound decisions.

\subsubsection{Control experiment: the disruption impairs outbound decisions even when task variables are observed}

    Finally, we performed a control simulation to identify more clearly why the disruption impairs the ability to make the outbound decisions. It is worth noting that in the higher-level (prefrontal) map of the HAI model, the task variables (color-coded nodes in Figure~\ref{fig:simulation1}) are hidden (i.e., they do not have corresponding observations), unlike the spatial states in the hippocampal map that have corresponding observations. Hence, the agent's estimate of the hidden state at the level of task space depends entirely on spatial memory, as encoded in the model's transition function -- which we disrupt.
    
To assess whether the partial observability of task space is key to explain the effects of the disruption, we considered an alternative (control) environment where the task variable is observable, however, the component required for spatial memory is unobserved. We consider the following variant of the W-maze environment: the maze is augmented with a light signal. Depending on where the agent just collected reward, a different color is shown. We implement this by adding an additional observation modality containing a categorical variable, indicating the previous rewarding corridor.

We repeated the same disruption experiment as before and found that the disrupted HAI agent maintains a high success rate for inbound decisions, but makes outbound decisions at chance level (Figure~\ref{fig:withlight}). This control experiment therefore indicates that disruption of spatial memory impairs the ability to make the outbound decisions, even when task variables are observable.
    
    \subsection{Experiment 2: learning multiple rules and flexible switching between them}

    In the next experiment, we investigate whether the HAI agent is able to learn multiple alternation rules as proposed in~\citep{den_bakker_sharp-wave_2022}. We consider three spatial alternation rules, where for each rule, a different corridor serves as the alternation point (LCLR, LCRC, and RCRL). Note that for simplicity, we use the same structure and notation as in the previous experiment with the W-maze, despite the study of multiple alternation rules of~\citep{den_bakker_sharp-wave_2022} has been conducted on a three-arm radial maze, not a W-maze.
    
    In this experiment, the HAI agent is endowed with a cognitive map of physical space which is the same as Experiment 1 (since the maze is always the same). However, it has to learn a new cognitive map of task space, which now comprises the dynamics of the three rules. For this, we let the agent explore the maze for 8000 steps, with alternating task rules after blocks of 1000 steps (in task space, i.e. visits to corridors). To aid the learning process, we guide the sequence and select the correct action 75\% of the time, and random otherwise.

  \begin{figure}[t!]
        \centering
        \begin{subfigure}{.45\textwidth}
            \centering
            \includegraphics[width=\textwidth]{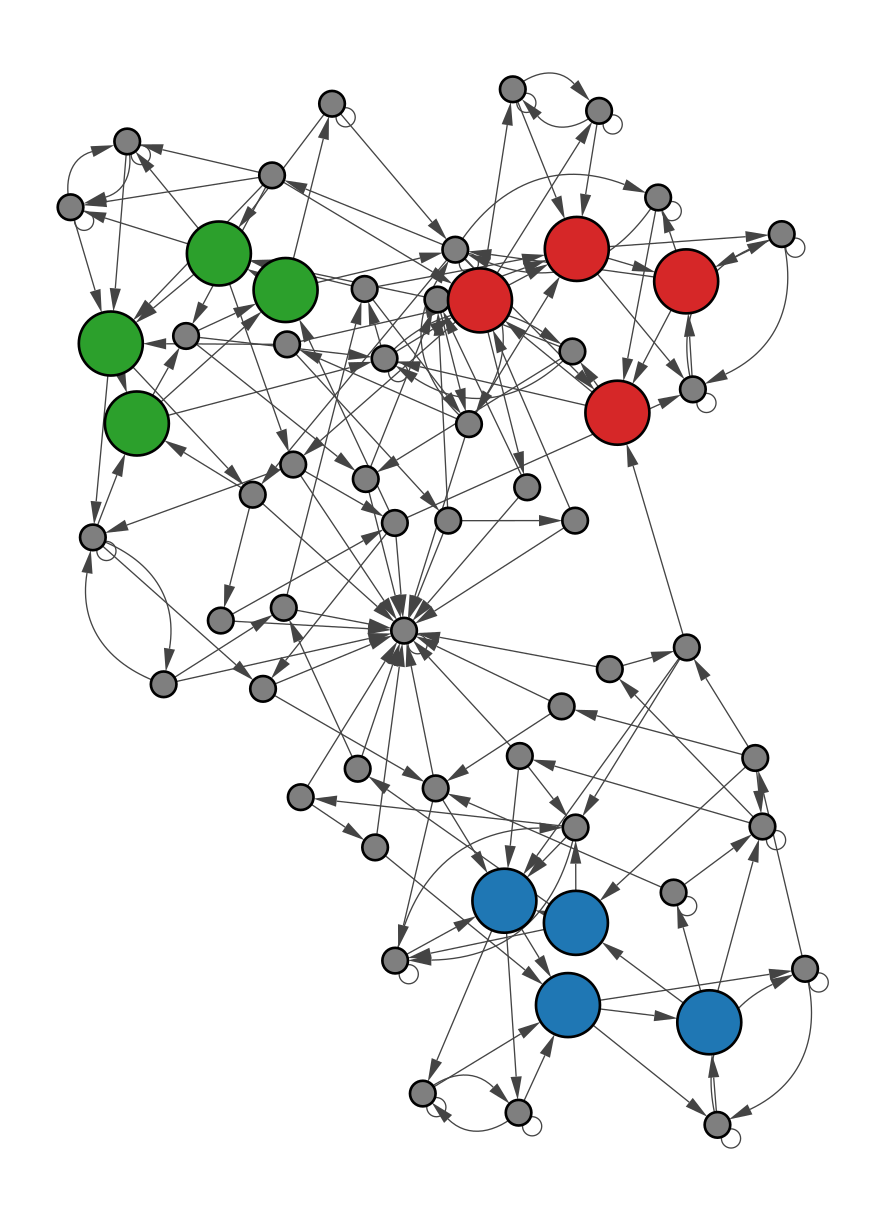}
            \subcaption[]{\label{fig:map_three_rules}}
        \end{subfigure}
        \hfill
        \begin{subfigure}{.38\textwidth}
            \centering
            \includegraphics[width=\textwidth]{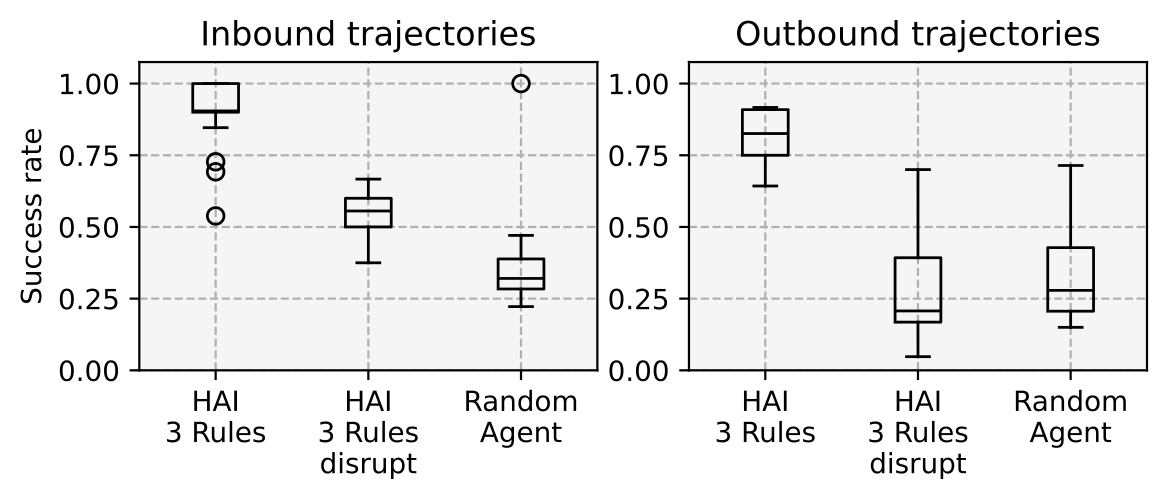}
            \subcaption[]{\label{fig:boxes}}
            
            \includegraphics[width=\textwidth]{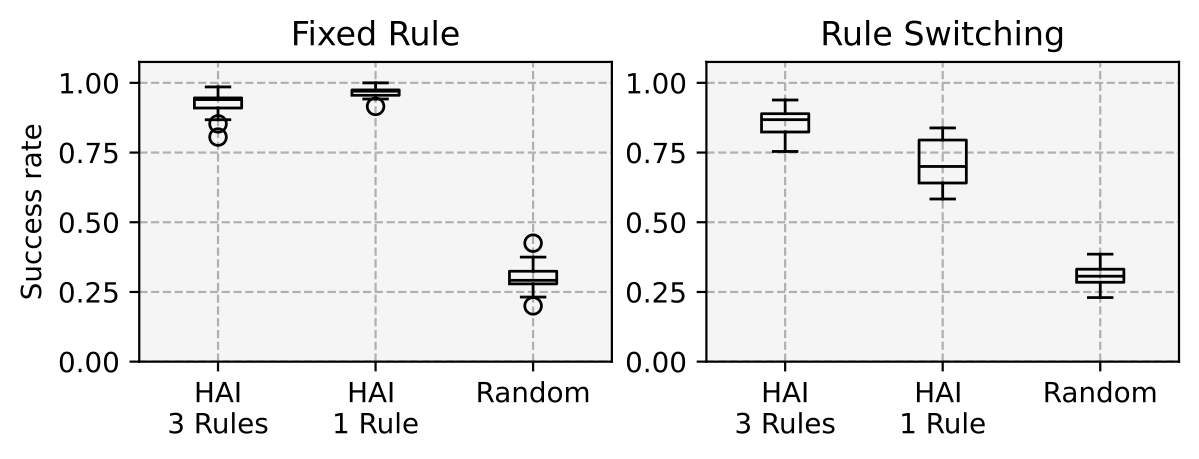}
            \subcaption[]{\label{fig:rule_switching_box}}
            
            \includegraphics[width=\textwidth]{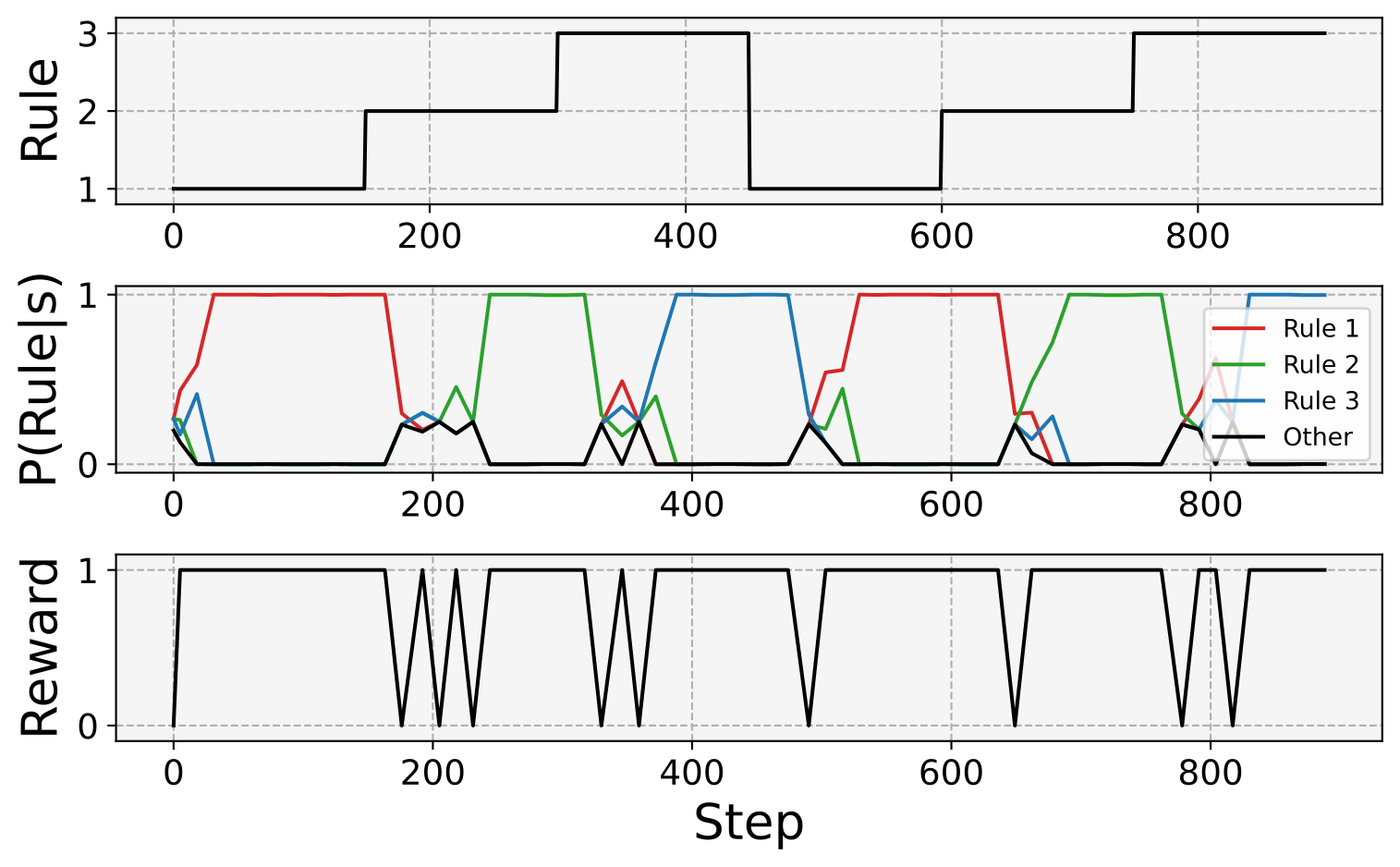}
            \subcaption[]{\label{fig:rule_switching}}
        \end{subfigure}
        \hfill
        \caption{\textbf{Experiment 2: Multiple spatial alternation rules (LCLR, LCRC, and RCRL).} (a) Representation of the states used in the CSCG for the three distinct rules. States again map to the conjunction of the visited corridor and the presence of reward. Colors in the graph indicate the rule for which the state is used (red for rule 1, green for rule 2, and blue for rule 3). The gray nodes represent other states, either without reward or in transition between multiple rules 
        (b) Disruption experiment for the model that learned three rules, evaluated on a single rule. The success rate is computed over 20 trials per agent, with 300 steps per trial.
        (c) Average reward for the agents trained on 3 rules, 1 rules and random selected. The case where the rule is constant (left), and the rule switches every 300 steps to a randomly selected next rule (right) is considered. This experiment was repeated over 20 trials of 900 steps. 
        (d) Performance during a single representative trial of 1000 steps, during which the rule switches randomly, every 150 steps. The top plot shows the rule that is currently in play. The center plot shows the belief over rule, which is computed by first extracting states belonging to a single rule, and then measuring the likelihood of being in one of these states. The bottom plot shows the rewards collected over time.  
        }
        
        \label{fig:simulation3}
    \end{figure}

    The learned cognitive map of task space is visualized in Figure~\ref{fig:map_three_rules}. For ease of visualization, we extract the states that are active when the agent has correctly inferred the rule, after a warmup period (15 steps) in each block. Similar to Section~\ref{sect:sim1}, the states of the task space correspond to the distinct colors, however since the different rules have distinct transition dynamics, these are mapped to unique states. Figure~\ref{fig:map_three_rules} uses different colors to visualize the states of the three different rules (red for rule 1, green for rule 2, and blue for rule 3). The smaller dark gray states are states used for transitioning between rules, or encode the corridors where no reward is found. We observe that for each of the individual rules, four states are active. This is an indication of correct learning since four is the optimal number required for each alternation problem. 

    Then, we evaluate whether the HAI agent is able to infer which rule is currently in place -- in order to continue collecting rewards when the rule switches. For this, we test the HAI agent (endowed with a preference to maximize reward) in a task in which the rule switches every 300 steps. 

    Figure~\ref{fig:boxes} shows the success rate of various agents for a fixed rule over 20 trials. From the success rate of the HAI agent trained on the three rules (HAI 3 rules), we can conclude that the agent properly encodes the rule. When we apply the same disruption as in Experiment 1 (Section~\ref{sect:sim1}), the success rate for inbound trajectories is now also affected (HAI 3 rules disrupt). This drop in performance is caused by the fact that when considering multiple rules, there is also ambiguity over which rule is currently in play, for which the spatial memory component is crucial.
    
    Figure~\ref{fig:rule_switching_box} compares the performance of various agents in two scenarios: in the former, there is a single rule, which remains fixed throughout the experiment (left plot), whereas in the latter, there are three rules, which switch randomly every 300 steps, over 20 trials. The left plot shows that both the Active Inference agents trained on three rules (HAI 3 rules) and on one rule (HAI 1 rule) achieve a very high success rate in the first scenario and both outperform a random agent (random). This result indicates that learning multiple rules does not decrease performance when a single rule is in place. The right plot shows that the Active Inference agent trained on multiple rules (HAI 3 rules) achieves a high success rate in the second scenario, too, indicating it correctly learned all the rules and successfully alternates them. The Active Inference agent trained on a single rule (HAI 1 rule) achieves a lower success rate, as expected, but still it outperforms a random agent. This result shows that learning only a subset of the task rules could lead to a relatively good performance in the W-maze, but still it is possible to discriminate between agents that learn the task completely, partially or select randomly from their behavior.
    

    Figure~\ref{fig:rule_switching} illustrates the behavior of the agent during a single, representative trial. As shown in the figure, shortly after the rule switches (top panel), the HAI agent is able to correctly update its belief about the current rule in its plan (center panel) and secure rewards (bottom panel). The inference of the rule currently in play follows standard Bayesian approach: when the rule changes, the agent receives unexpected observations (since expected rewards are not delivered) and correctly makes a transition to the subspace of the task rule map that encodes the most likely rule. In the center panel of Figure~\ref{fig:rule_switching}, we visualize the belief over each rule as the probability of being in one of the four states used in a particular rule (the rules are color-coded as Figure~\ref{fig:map_three_rules}) in or in another phase if the state does not occur in either of the rules (grey line). When the agent has a high probability of a rule, it no longer misses the reward. In some cases, it misses the reward because there is a potential overlap between the two rules. For example, around step 200 it could be either of the rules, as it only picks the right one once, and the wrong one after -- which could happen for each corridor in each of the rules. Note that in this architecture, the belief about the currently active rule is implicit in the model, but it could be made explicit by adding a hierarchical layer that maintains a probability distribution over the map or the task one is currently in, as in~\citep{stoianov_hippocampal_2022}.
    
    \subsubsection{Disrupting spatial memory yields specific patterns}
    \label{sect:sim3}

    Another important finding of the study of~\cite{den_bakker_sharp-wave_2022} is that disrupting optogenetically the medial prefrontal cortex (mPFC) immediately after a hippocampal SWR is detected significantly impairs performance, by increasing the occurrence of three distinct maladaptive patterns: (i) rotated alternation, where the agent follows a different rule, (ii) back and forth, where the agent alternates between two corridors, and (iii) circling behavior, where the agent iterates over all corridors in a cyclic pattern (see Figure~\ref{fig:error_patterns} for a graphical illustration of these patterns). The authors reported a proportion of 35\% alternation, 20 \% rotated alternation, 15 \% back and forth, and 10 \% circling.
    
    The behavioral patterns that are generated from the same type of disruption as conducted in experiment 1 - where the transition model of the POMDP is lesioned - are investigated. We match the patterns observed in the biological counterpart of this experiment~\citep{den_bakker_sharp-wave_2022}, and evaluate the trajectories generated by the agent trained on the three rules, for an agent with and without the spatial memory disruption from Section~\ref{sect:sim1}, as well as a random agent. The results are reported in Table~\ref{tab:disruptionpatterns}. This shows that, when lesioned, our model reproduces all four behavioral patterns seen in lesioned animals~\citep{den_bakker_sharp-wave_2022} (alternation, rotated alternation, back and forth, and circling) although in slightly different proportions compared to the empirical study (please see Appendix~\ref{app:gamma} for a simulation how the choice of $\gamma$ affects these proportions). We observe that for the disrupted agent, a large part of the behaviors can be classified as back-and-forth behavior. The specific proportions of categorization are a function of the agents' habit policy E. This habit acts as a prior over the selected action (of level 2). When no temporal information is considered through the lesion, this prior will dominate, and the agent will act according to the statistics of the training data. This can be observed by looking at HAI disrupt (1), where the agent behavior falls back to either the alternation or the back-and-forth case. Finally, we also note that the patterns generated by the random agent are generally classified as the \emph{other} category, indicating that the behavior generated by our disruption is not random. 
    
    \begin{table}[]
        \centering
        \caption{\textbf{Behavioral patterns during disruption:} Classified observed behavioral patterns according to~\cite{den_bakker_sharp-wave_2022} in different scenarios for agents trained on the three rules indicated by (3) and a single rule indicated by (1).}
        \label{tab:disruptionpatterns}
        \begin{tabular}{l|c|c|c|c|c}
            & \textbf{HAI} & \textbf{HAI} & \textbf{HAI} & \textbf{HAI }  & \textbf{Random}\\
            & \textbf{control (1)} & \textbf{disrupt (1)} & \textbf{control (3)} & \textbf{disrupt (3)} & \\
            \hline
            Alternation        & 86.33\% & 49.07\% &  76.00\% & 16.80\% &  3.47\%   \\
            Rotated alternation& 0.00\%  & 0.80\%  &  1.07 \% & 13.55\% &  8.24\%  \\
            Back and forth     & 2.68\%  & 38.40\% &  2.40 \% & 38.48\% &  8.86\%   \\
            Circling           & 1.61\%  & 1.60\%  &  8.80 \% & 17.34\% &  5.86\%  \\
            Other              & 9.38\%  & 10.13\%  &  11.73\% & 13.82\% &  73.75\% \\
        \end{tabular}
    \end{table}

    Furthermore, we performed the same experiment for an agent trained on a single rule. We observe that both rotated alternation and circling behavior drops to near 0\% as shown in the table indicated by HAI control (1) and HAI disrupt (1). This is due to the agent not considering other rules, when falling back to its habit policy. To the best of our knowledge, the pattern of errors in animals trained with a single rule have not been systematically studied and hence our results could be considered as a novel prediction of the HAI model.

    \subsubsection{Disrupting communication between the two levels yields perseveration behavior}
    \label{sec:disrupting_communication}
    
    The previous simulations illustrate that disruptions of the HAI model permit reproducing empirical findings of rodent studies~\citep{den_bakker_sharp-wave_2022,jadhav_awake_2012}. However, our model permits realizing synthetic experiments -- for example, other types of disruptions -- that have not been studied empirically so far, but which could be interpreted as novel predictions of the model. 
    
\begin{figure}
        \centering
        \includegraphics[width=0.4\textwidth]{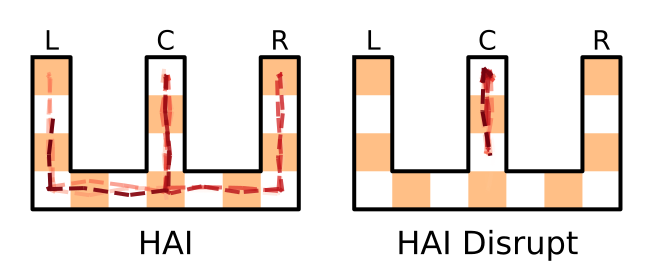}
        \caption{\textbf{Disruption experiment for rule switching}: when communication between the level 1 and level 2 model is suppressed, perseveration behavior is observed. Trajectory order is indicated by the shade where lighter is earlier, noise was added to the trajectory to visualize overlapping visits.}
        \label{fig:perseveration}
    \end{figure}

    Here, we investigate the potential effects of a disruption of hippocampal - prefrontal communication that completely impairs the communication between the two levels of the HAI model. Specifically, in this simulation, we prevent level 1 from sending any bottom-up information to level 2. This is a severe impairment, since hierarchical inference rests upon the reciprocal messages passing between the two levels -- and level 2 only receives observations from level 1, not from the environment. Figure~\ref{fig:perseveration} shows the differences between the physical trajectory of the intact HAI agent (left) and the disrupted HAI agent (right) in which communication between the two levels is impaired. The shade of the trajectory indicates the order of visits (lighter is earlier). The figure shows that while the intact HAI agent follows the correct rule, the disrupted HAI agent perseveres in choosing the center corridor. 

    Interestingly, the agent does check at the unambiguous T-junction to self-localize and ensure that it is still in the correct corridor (which is a feature of the epistemic dynamics of active inference, see Section~\ref{sect:hai}), but never selects the next goal in the task rule. This is because, without the bottom-up message from level 1, level 2 is unable to correctly recognize that one phase of the task has been achieved and hence never updates the top-down plan. This result illustrates the possibility to use the model to generate novel predictions -- in this case, about how disrupting communication between the two levels produces perseveration -- which could be subsequently tested empirically.
 
\section{Discussion}

Here, we advanced a novel computational theory of hippocampal (HC) - prefrontal (PFC) interactions during cognitive tasks that require navigating in both physical and task space -- such as spatial alternation tasks. Empirical studies of spatial alternation have assessed that they depend on the animal's spatial memory, which is at least in part maintained by HC-PFC communication \citep{jones_theta_2005,spiers_keeping_2008,shin_multiple_2016,patai_versatile_2021,tang_multiple_2021,simons2003prefrontal} and that disruption of this communication can impair difficult (outbound) decisions~\citep{jadhav_awake_2012} and the ability to switch between multiple rules \citep{den_bakker_sharp-wave_2022}. 

Our computational model is based on -- and unites -- two established theories. The former is a theory of cognitive map formation, based on a statistical sequence learning algorithm: the clone-structured cognitive graph (CSCG) \citep{george_clone-structured_2021}. Previous studies assessed that CSCG are computationally effective and have biological validity since they are able to successfully reproduce a number of empirical observations about cognitive map formation in the hippocampus, including for example the emergence of place cell coding and splitter cells~\citep{raju_space_2022,sun_learning_2023} and orthogonalized state representations~\citep{sun_learning_2023} (see also~\citep{whittington_tolman-eichenbaum_2020,whittington_how_2022} for alternative proposals about the computational principles that underlie spatial map formation). Here we extended this body of work by showing that CSCGs can be used in a hierarchical scheme to learn not just cognitive maps for physical space (putatively linked to hippocampal computations), but also more abstract cognitive maps for task space (putatively linked to prefrontal computations). Furthermore, we used learned CSCGs as components of a hierarchical generative model for active inference \citep{parr_active_2022}. Active inference is a normative framework to model sentient behavior in terms of free energy minimization and (approximate Bayesian) inference over a generative model, which is gaining popularity in cognitive neuroscience. We have shown that by combining two learned CSCG maps (for physical and task space), it is possible to design an effective hierarchical active inference (HAI) agent able to solve spatial alternation tasks. Interestingly, this scheme affords (hierarchical) planning by only using local computations -- that is, top-down and bottom-up message-passing between the two hierarchical levels. 



Furthermore, by simulating the interruption of HC - PFC communication in our model, we were able to correctly reproduce impairments of difficult (outbound) decisions~\citep{jadhav_awake_2012} (Experiment 1) and of correct rule switching, unveiling the same kind of maladaptive behavior -- rotated alternation, back and forth, and circling behavior (Experiment 2) -- observed empirically in rodents \citep{den_bakker_sharp-wave_2022}.

Our model suggests that the selective impairment of outbound decisions provoked by hippocampal SWR disruptions~\citep{jadhav_awake_2012} is due to the fact that the SWRs convey messages to higher structures, like the PFC, which are used to update a belief about the current stage of the task (specifically, this message is key to propagate the belief about task state at level 2 over time). This interpretation is in keeping with the idea of~\citep{jadhav_awake_2012} that the impairment is at the level of spatial memory, not of hippocampal place coding, but to our knowledge, ours is the first work that provides a mechanistic model of this theoretical proposal. Furthermore, our model suggests that the maladaptive behavior found in \citep{den_bakker_sharp-wave_2022} could be due to the impossibility for the higher, prefrontal component (level 2) to correctly update its belief, based on bottom-up message passing from the hippocampal component (level 1). This perspective is coherent with the finding of \citep{den_bakker_sharp-wave_2022} that the only disruption of mPFC that prevents spatial rule switching is one that directly follows hippocampal SWRs -- hence highlighting the importance of coherent HC - PFC reactivations to solve spatial alternation tasks. Finally, our simulations suggest that other, more severe interruptions of HC - PFC communication (Section~\ref{sec:disrupting_communication}) could produce a specific pattern of maladaptive behavior -- namely, perseverative behavior -- that differs from those observed in the above disruption experiments. This is a novel prediction that remains to be tested empirically.

Besides helping understand HC - PFC interactions, our simulations (Appendix \ref{sec:noisy_scelario}) suggest that looking at the animal's epistemic behavior (e.g., the selection of actions to self-localize and reduce uncertainty about the current pose) during uncertain tasks could be important to elucidate its strategy; in particular, whether it aims to maximize reward or also to minimize its uncertainty -- as predicted by active inference \citep{parr_active_2022,schwartenbeck_computational_2019,rens2023evidence}. Future studies might assess whether epistemic imperatives are important drivers of behavior during spatial alternation or similar tasks, especially in conditions of high uncertainty, such as when the animal is randomly placed in a maze (as in our simulations) or when spatial contingencies or task rules change unexpectedly.

It could, in principle, be possible to solve the navigation tasks studied in this article using a non-hierarchical generative model with a single ``map'' (and a single CSCG) that encompasses both spatial and task-related components. However, the hierarchical structure of the HAI generative model used in this study better reflects the implicit division of labor between HC and PFC circuits, which is well established empirically in rodent studies. For example, inactivating prefrontal structures during navigation tasks tends to disrupt rule-related contingencies and deliberative behavior, while sparing spatial representation \cite{den_bakker_sharp-wave_2022,schmidt2019disrupting}. Therefore, our model reflects the widespread perspective that goal-directed navigation depends on the coordinated interplay between (inferential) processes at two levels, which could be associated with HC and PFC structures \citep{shin_multiple_2016,tang_multiple_2021,patai_versatile_2021,ito2015prefrontal}. The division of labor between HC and PFC is also central to other prominent accounts, such as the Complementary Learning Systems framework, which highlights that faster learning in HC facilitates slower learning in neocortex -- with the latter integrating across episodes to extract semantic structure \citep{mcclelland1995there}. Furthermore, from a computational perspective, the hierarchical structure of the HAI model affords a useful factorization: learning a novel rule in a known maze, as we did in Experiment 2, only requires re-train the higher-level CSCG, while leaving the lower-level CSCG unchanged. This might be more problematic when using a single CSCG that combines spatial and task-related information.

The current study has several limitations that will need to be addressed in future studies. First, for efficiency reasons, we learned the CSCGs offline (before embedding them into the HAI), using a simplified procedure: we used predefined trajectories that exhaustively covered the W-maze as inputs for the cognitive map of physical space and 75\% correct trajectories as inputs for the cognitive map of task space. In the future, it would be interesting to train CSCG online, similar to~\citep{lazaro-gredilla_fast_2023}, by guiding the exploration through active inference dynamics~\citep{friston_active_2017,schwartenbeck_computational_2019,parr_active_2022}. A second research avenue is to relax the separation of the timescales between the two levels, by selecting their inputs (e.g., level 1 takes all sensory observations as inputs, whereas level 2 only considers observations that could be rewarding -- and in particular, observation 1 in Figure \ref{fig:observations}). In the future, it would be interesting to explore methods to learn hierarchical models with multiple timescales~\citep{yamashita2008emergence,hinton2006fast} and effective state spaces for navigation and for task rules in self-supervised (and/or reward-guided) ways, as shown in prior work~\citep{stoianov_hippocampal_2022,stoianov2018model,stoianov2016prefrontal,niv2019learning}. This might also help understand the reciprocal influences between cognitive map learning at different levels and in different (e.g., prefrontal versus hippocampal) brain structures. A third challenge is to avoid having the agent learn from scratch each new maze or rule. Recent work in transfer learning shows that it is possible to reuse existing cognitive maps or latent task representations to learn novel and similar tasks much faster~\citep{stoianov_hippocampal_2022,guntupalli2023graph,stoianov2016prefrontal,swaminathan_schema-learning_2023}. Extending our architecture with transfer learning abilities would be important to provide more accurate models of how animals learn cognitive maps, especially given the strong evidence for the reuse of existing neural sequences and cognitive maps in the hippocampus~\citep{liu2019preconfigured,farzanfar_cognitive_2023}. 

Another limitation of the model presented in this study is that in the CSCG of the task layer, we made explicit the fact that the relevant task states are the ends of corridors, because the animals can (only) acquire rewards in such states. This design choice reflects the animals' knowledge, given that in the spatial alternation task \citep{jadhav_awake_2012}, the three arms had reward food wells at their endpoints. However, future work should consider generative models that learn autonomously what the relevant task states are, even if they are not explicitly cued as in \cite{jadhav_awake_2012}. There is an established literature showing that latent task states relevant for cognitive control can be learned autonomously using Bayesian (nonparametric) methods \citep{collins2013cognitive,stoianov2016prefrontal,stoianov2018model,gershman2012tutorial} and that task rules can be learned using deep reinforcement learning \citep{wang2018prefrontal}. However, it remains to be assessed in future research how to integrate these methods in the HAI model.

Future studies might consider how to adapt the HAI agent to robot navigation and planning. There is increasing interest in applying active inference to robotic domains, given the appeal of its unified approach to control, state estimation, and world model learning~\citep{lanillos_active_2021,da_costa_how_2022,taniguchi_world_2023}. Hierarchical active inference is especially appealing, since it facilitates planning over long time horizons. Indeed, the computational burden required to calculate the expected free energy of long policies is high, but it can be substantially reduced by splitting work between (higher-level) policies that select (a sequence of) subgoal(s) along with (lower-level) policies that select the specific actions to reach each subgoal \citep{donnarumma2016problem}. For example, one study built a hierarchical model for robot navigation, using multiple layers of recurrent state space models~\citep{catal_robot_2021}. Another study realized a hierarchical model for next-frame video prediction, using a  subjective timescale for the predictions ~\citep{zakharov_variational_2021}. In another study~\citep{van_de_maele_integrating_2023}, a single layer CSCG was embedded within the active inference framework and was able to support navigation in different mazes. An interesting direction of future work could be adding a (learned) hierarchical layer that maps the high-dimensional observations space to a discrete state space and stack the proposed hierarchical active inference on top with multiple layers of CSCGs. This would potentially afford abstract reasoning and planning for complex tasks, directly from sensory observations such as pixels.

Finally, future studies might consider in more detail the functional role and content of hippocampal reactivations and replay in the hippocampus and other brain structures, such as the prefrontal cortex and the ventral striatum \citep{wilson_reactivation_1994,foster2017replay,pfeiffer_hippocampal_2013,liu_human_2019,lansink2009hippocampus,peyrache2009replay,wittkuhn2021dynamics,liu2018generative,buzsaki2015hippocampal,buzsaki2019brain,gupta2010hippocampal,nour2021impaired,pezzulo2014internally,pezzulo2017internally}. Previous studies suggested that replay might play different roles, which range from memory functions to planning, compositional computation and the optimization of the brain's generative models~\citep{stoianov_hippocampal_2022,shin_continual_2017,mattar2018prioritized,pezzulo2019planning,pezzulo2021secret,wittkuhn_replay_2021,kurth2023replay}. However, these works have mostly considered hippocampal replay, not coordinated replay in the hippocampus and the prefrontal cortex (and other brain areas). It might be interesting to combine the insights of these studies with the hierarchical architecture of the HAI agent, to test whether (for example) planning and model optimization benefit from the combined replay of multiple brain structures, as opposed to local replay in the hippocampus.



\section{Methods}
    \label{sect:method}

    In this paper, we develop a hierarchical planning agent by combining two components: we use \emph{clone structured cognitive graphs (CSCG)} to learn ``cognitive maps'' of physical and task space; and \emph{hierarchical active inference} to form hierarchical plans. In the next sections, we explain the two key components of our approach: CSCG (Section~\ref{sect:cscg}) and hierarchical active inference (Sec.~\ref{sect:hai}). Finally, we explain how we combined these (Section~\ref{sect:map}).
    
    \subsection{Clone-structured cognitive graphs (CSCG) for learning cognitive maps of physical and task space}
    \label{sect:cscg}

    \begin{figure}
        \centering
        \begin{subfigure}{0.49\textwidth}
            \includegraphics[width=0.80\textwidth]{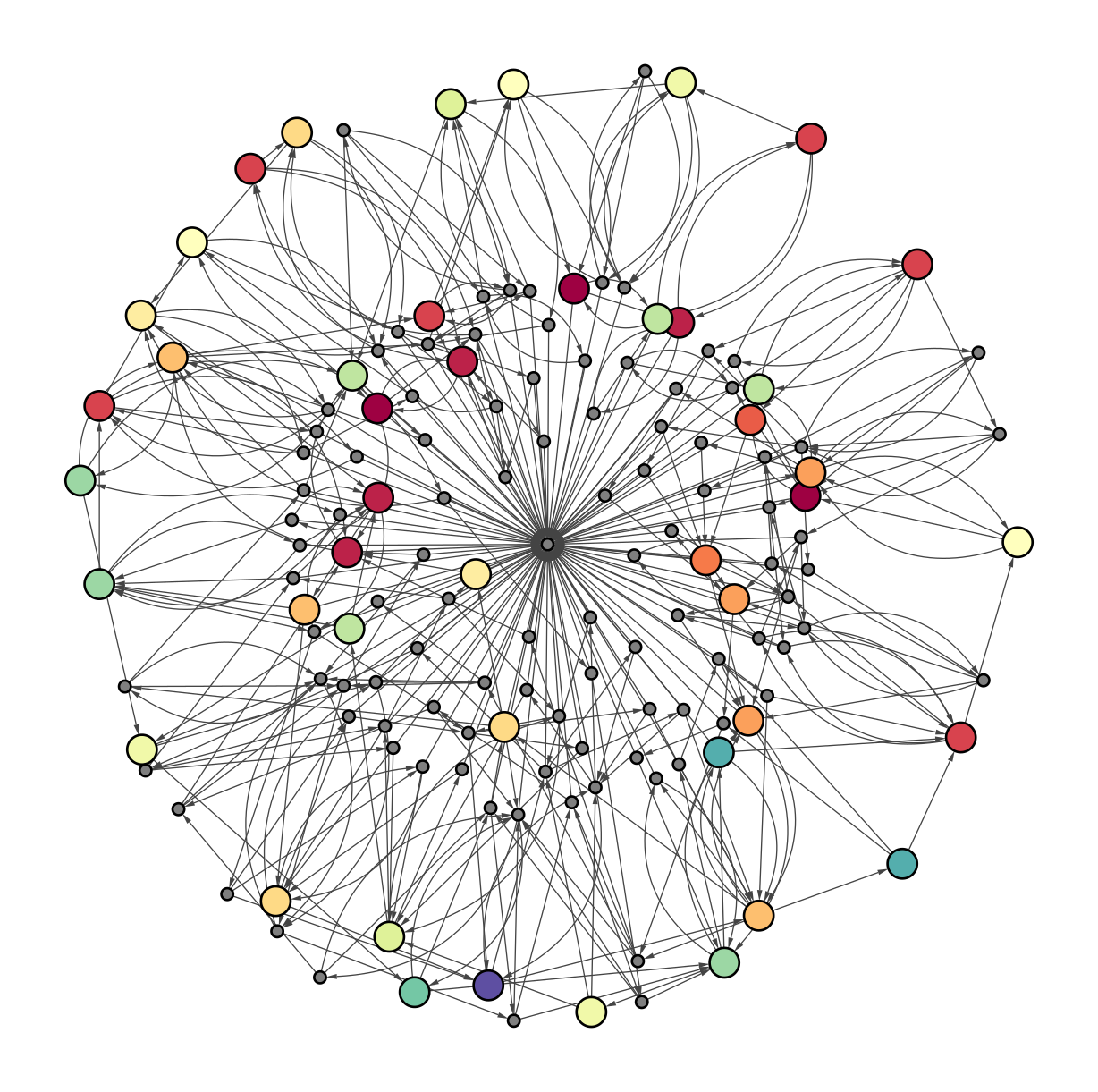} 
            \subcaption[]{\label{fig:level1graph}}
        \end{subfigure}
        \hfill
        \begin{subfigure}{0.49\textwidth}
            \includegraphics[width=0.80\textwidth]{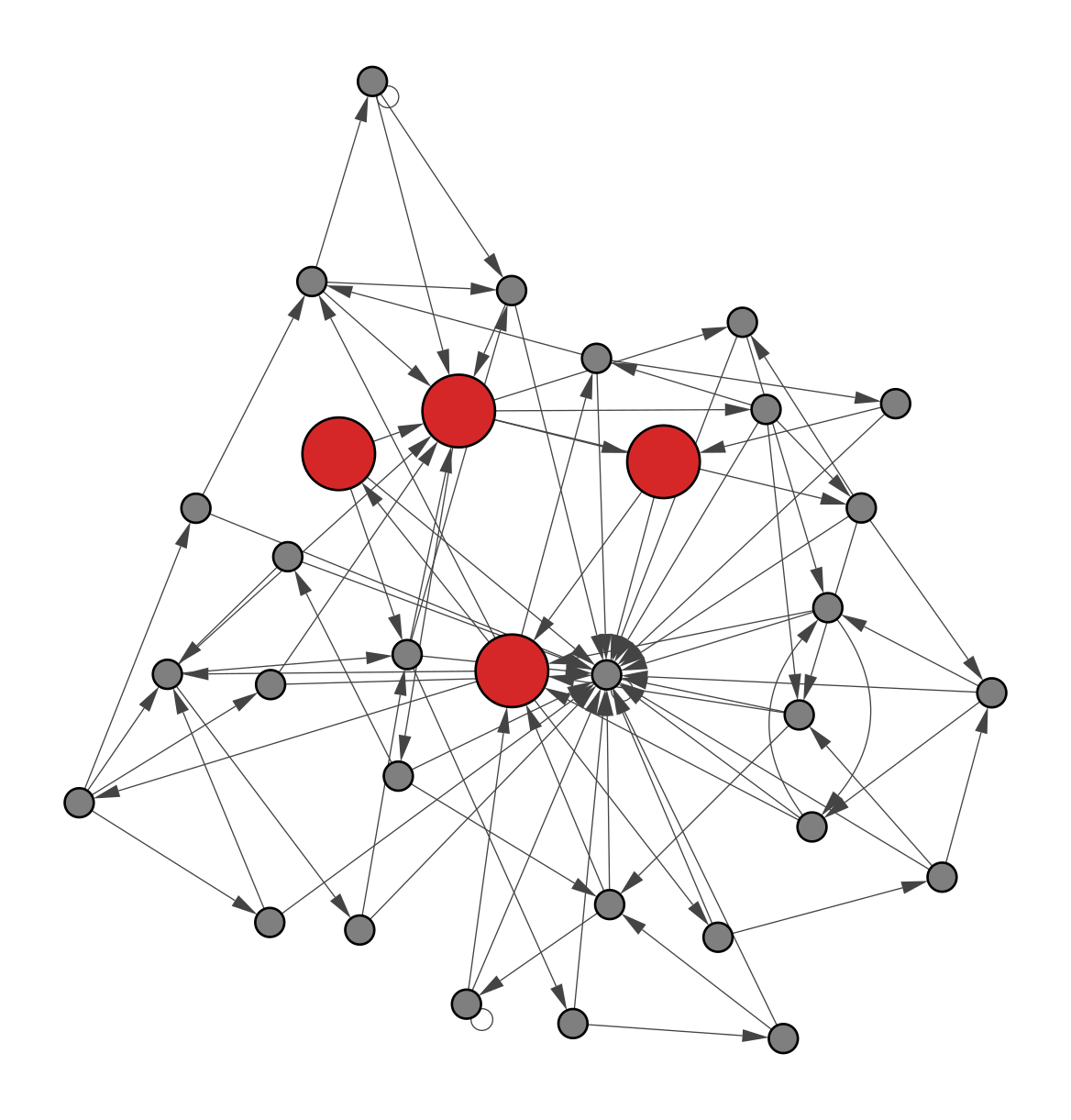} 
            \subcaption[]{\label{fig:level2graph}}
        \end{subfigure}
        \caption{\textbf{Learned cognitive graphs.} The arrows indicate possible transitions from one state to the next. The node in the center where all nodes point to, is the added and absorbing ``dummy'' state.  (a) The learned CSCG for the physical space model (level 1). The color in the graph represents distinct observations, only the states that are active when pursuing the spatial alternation rule (LCRC) are colored, the other states are marked in gray. (b) The learned CSCG for the task space model (level 2). The color in the graph indicates the states that were active when pursuing the spatial alternation rule (LCLR). While only four states are necessary to encode the rule, it can be observed that many more states are added by learning the other paths that might have been visited during training.}
        \label{fig:learned_graphs}
    \end{figure}

    Clone-structured cognitive graphs (CSCGs) are a probabilistic model for representing sequences of data, e.g. a sequence of action-observation pairs~\citep{gothoskar2019different,george_clone-structured_2021}. They are a special case of Hidden Markov Models (HMM), where each observation maps to a subset of the hidden states: the so-called clones of this observation. While these states have the same observation likelihood, they differ in the implied dynamics encoded in their transition model. Through the sequence of action-observation pairs, specific clones will have a higher likelihood and can therefore disambiguate the aliased observations. 

    The CSCGs are learned using the expectation-maximization (EM) algorithm for HMMs (the Baum-Welch algorithm) which maximizes the ELBO as described in~\cite{george_clone-structured_2021}. During the E-step, the posteriors over states are estimated through smoothing, i.e. the forward-backward algorithm. The M-step then selects the optimal parameters for the transition model, given this sequence of visited states. For the update equations, the reader is referred to~\cite{george_clone-structured_2021}.  

    We consider two cognitive maps, that represent the W-maze task at two distinct levels. The first level considers the structure of the environment (i.e. where can the agent walk and where are the walls), and the second level encodes the task rule (i.e. in which order the corridors yields the reward). We learn these two maps in a sequential fashion, using two distinct CSCGs. 
    
    We first collect a sequence of data for learning the spatial structure by exploring the maze. We designed the exploration sequence to exhaustively cover the W-maze such that a path from each pose to each possible other pose is present. This sequence was used to learn the parameters of a CSCG for the first (physical) level, using the expectation-maximization mechanism described in~\citep{george_clone-structured_2021}. We initialize the level 1 CSCG with 20 clone states per observation. After learning, the model is pruned using a Viterbi-decoding algorithm as described in~\citep{george_clone-structured_2021} and mapped to the POMDP of the first level of the hierarchical generative model (as will be explained in Section~\ref{sect:map}) for engaging in active inference. The learned graph is visualized in Figure~\ref{fig:level1graph}. The nodes in the graph correspond to the learned nodes and their colors represent the observations encountered in the states, during the correct execution of the spatial alternation task. The gray nodes represent other states required for transitioning between states or trajectories outside the correct spatial alternation. Note that the graph shown in Figure \ref{fig:dynamics} only shows the colored nodes, but not the gray nodes.

    We use the learned cognitive map of physical space to learn the cognitive map of task space. We first extract the states that are distinctly representing the end of each corridor (i.e. matching with observation 1 in Figure~\ref{fig:observations}) and use them as the actions for the task level. When an action is selected, this means that the agent is moving toward one of these states, and thus the end of one of the corridors. The observations at this level are the presence of reward (``1'' if the agent is following the rule, ``0'' otherwise), and the reached level 1 state (physical level). To aid the learning process, we guide the agent to follow the rule 75\% of the time. We now learn a CSCG with 10 clones per observation using the same expectation-maximization scheme for the single rule scheme, and 32 clones for the scenario with three rules. After learning, the model is pruned using a Viterbi algorithm as described in~\citep{george_clone-structured_2021} and mapped to the POMDP of the second level of the hierarchical generative model to engage in active inference (see Section~\ref{sect:map}). For a robustness analysis on the model capacity with respect to amount of rules, rule length and amount of clones per state, the reader is referred to the appendix. We visualize the learned cognitive graph in Figure~\ref{fig:level2graph}, where the states used when pursuing the rule are colored in red. It can be observed that the spatial alternation rule is encoded in this graph, using four states, as this is the optimal amount required for the spatial alternation rule (two states for the center, one for the left, and one for the right corridor). Note that the graph shown in Figure \ref{fig:dynamics} only shows the colored nodes, but not the gray nodes. 
    
    \subsection{Hierarchical active inference} 
    \label{sect:hai}
    
    \begin{figure}[t!]
        \centering
        \includegraphics[width=0.80\textwidth]{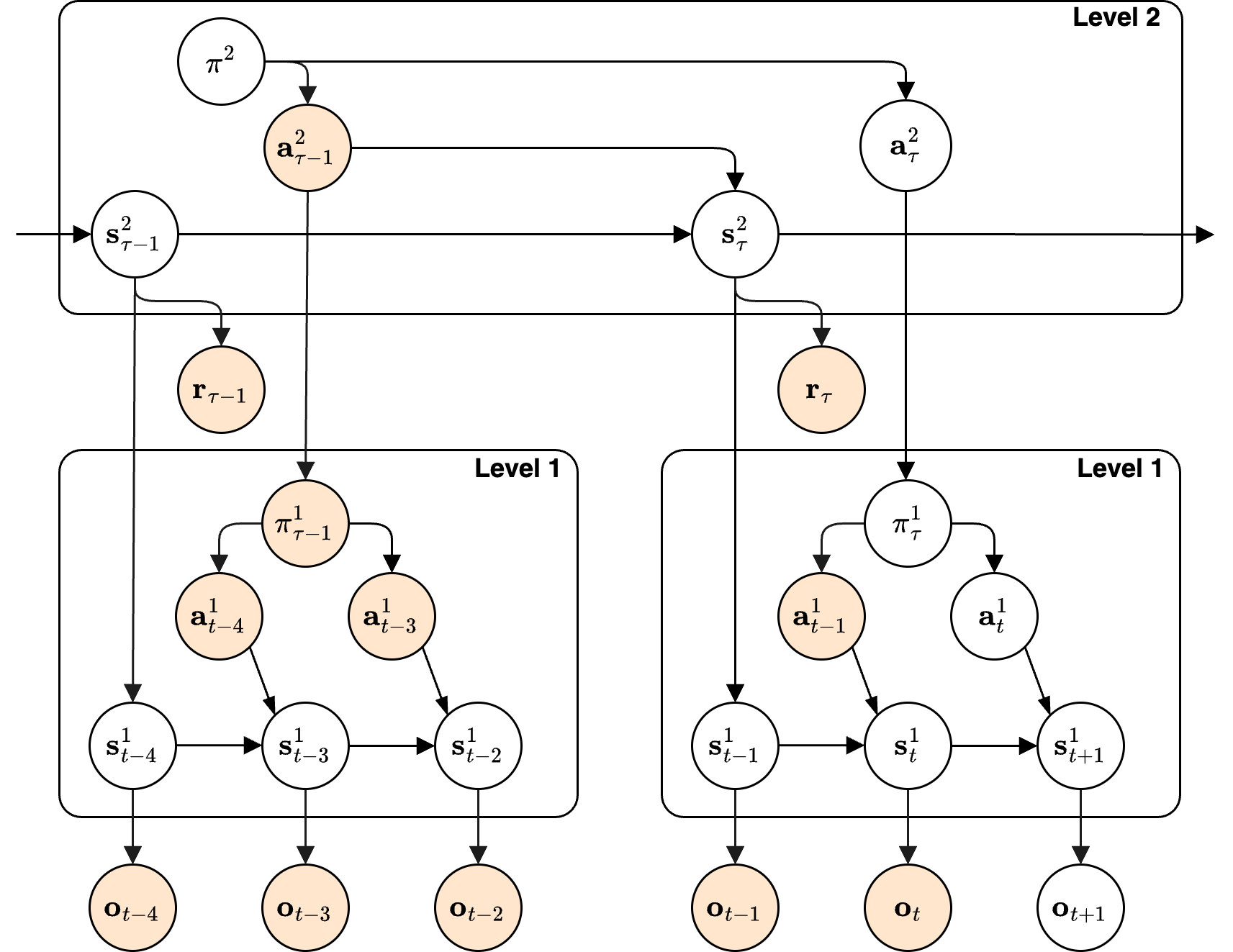}
        \caption{\textbf{The Hierarchical Generative Model:} This generative model consists of two hierarchical levels, where the top level operates at a slower timescale than the lower level. The policy at the highest level sets the actions for the top level, which determines temporal transitions between states at level 2. The level 2 state then generates the presence of reward and the level 1 state. At level 1, the states generate the observations and the policy generates the actions, depending upon the selected level 2 action. Finally, when a policy is pursued, a message is passed to the upper level, transitioning the upper level to the next state. This highlights the different timescales, whereas level 1 operates in the timescale of the agent's movement, the upper level operates at the timescale of the reached subgoals (visited corridors). Orange circles denote observed variables, while white circles denote unobserved variables.}
        \label{fig:generativemodel}
    \end{figure}

        
        Active inference is a normative framework to describe cognitive processing and brain dynamics in living organisms \citep{parr_active_2022}. It assumes that action and perception both minimize a common functional: the minimization of variational free energy (which is an upper bound to the organisms' surprise). Central to active inference is the idea that living organisms are equipped with generative models, capturing the causal relations between observable outcomes, the agent's actions, and hidden states. 

        It is important to note the difference between the agent's generative model and the generative process. The former represents the internal generative model that the agent uses to attribute consequences to its actions, while the latter represents the true process from which outcomes are generated in the world. Crucially, the agent is unable to observe the hidden state of the generative process, as they are separated through a Markov blanket given the observation and action hidden variables. However, the agent is able to perform actions and observe the generated outcomes \citep{parr_active_2022,pezzulo2018hierarchical,pezzulo2015active}.
    
        \subsubsection{The hierarchical generative model}

        We endow the active inference agent with the hierarchical generative model depicted in Figure~\ref{fig:generativemodel}. In this section, we provide a high-level overview of the generative model, whereas in Section~\ref{sect:map} we discuss implementation details.

        The hierarchical generative model illustrated in Figure~\ref{fig:generativemodel} is split into two distinct levels. The highest hierarchical level (level 2) reasons at a more abstract level, e.g. about which corridor to visit, while the lowest level (level 1) considers the step-by-step navigation in the environment. Each of these levels can be interpreted as an individual partially observable Markov decision process (POMDPs)~\citep{kaelbling1998planning}, operating at different timescales: faster for level 1 and slower for level 2. However, crucially, the two levels interact reciprocally by exchanging messages, as described below. 

        The hierarchical model supports the hierarchical selection of abstract policies that move the agent from one goal to the other, following the task rule(s) (at level 2) and from one spatial location to the other, based on the spatial map (at level 1). Conceptually, this hierarchical selection starts when the agent's preferred observation is set to find the reward (technically, in active inference, this is done by assuming a strong prior for the reward observation) and it involves both levels.
                
        The most fundamental goal of Level 2 is to select a policy $\pi^2$ that moves the agent through task space, by steering abstract actions $\mathbf{a}^{2}_{t}$ that follow the rule learned by the agent (or in the case of multiple rules, as in Experiment 2, the rule currently inferred to be in place). These actions indicate the location that the agent will then try to reach in the spatial map (level 1). In this way, the level 2-action conditions the level 1 policy $\pi^1$. The level 2 state $\mathbf{s}^2_{t}$ is a latent representation of the corridor in which the agent is currently located in, and where it comes from (i.e. where the agent is in rule space). This abstraction is enforced by the choice of considering the reachable states to be matched to observation 1 (Figure~\ref{fig:observations}), reflecting the fact that the agent knows the potential reward locations (which, in the animal experiments, are baited with food wells). The agent only considers reward on level 2 and thus generates the presence of reward directly from the level 2 state.

        At level 1, the policy $\pi^1$ is conditioned by level 2, by setting the preferred state as reaching one of the spatial locations encoded in the cognitive graph. The policy $\pi^1$ generates the low-level actions that navigate the agent from one pose (position and orientation, encoded in hidden state $\mathbf{s}_t^1$) to the next. Then, the agent receives observations $\mathbf{o}_t$ from the environment, which reflect its current pose in the maze and which permit updating the hidden state estimate. Note that reaching the preferred spatial location requires making multiple transitions at level 1, but level 2 only makes a transition between states when the preference at level 1 is reached (or equivalently when the level 2 action is ``performed''). This creates a separation of timescales because multiple transitions at level 1 are nested between two subsequent actions at level 2.

        \subsubsection{Active inference}
        \label{sect:ai}

        As shown in Figure~\ref{fig:generativemodel}, the generative model considered in this paper is a hierarchically stacked (two levels) POMDP. In order to reduce complexity, we first discuss active inference for a single layer POMDP, as depicted in Figure~\ref{fig:factorgraph}. At time step $t$, we have observation $\mathbf{o}_t$, action $\mathbf{a}_t$ generated from policy $\pi$, and state $\mathbf{s}_t$. The generative model is factorized as: 
        \begin{equation}
            P(\mathbf{\tilde{s}}, \mathbf{\tilde{a}}, \mathbf{\tilde{o}}) = P(\mathbf{s}_0) \prod_{t=1} P(\mathbf{o}_t | \mathbf{s}_t) P(\mathbf{s}_t | \mathbf{s}_{t-1}, \mathbf{a}_{t-1}) P(\mathbf{a}_{t-1}),
        \end{equation}

        where the tilde represents a sequence over time. As computing the posterior distribution over the state is intractable for large state spaces, we resort to variational inference and introduce the approximate posterior $Q(\mathbf{\tilde{s}}|\mathbf{\tilde{o}}, \mathbf{\tilde{a}})$, the variational free energy measures the discrepancy between the joint distribution and the approximate posterior:  

        \begin{equation}
            F = \mathbb{E}_{Q(\mathbf{\tilde{s}}|\mathbf{\tilde{o}},\mathbf{\tilde{a}})}[
                    \log Q(\mathbf{\tilde{s}}|\mathbf{\tilde{o}},\mathbf{\tilde{a}})
                    - \log P(\mathbf{\tilde{s}}, \mathbf{\tilde{a}}, \mathbf{\tilde{o}})
            ] \\
        \end{equation}

        Active inference agents minimize this quantity through learning (by optimizing the model parameters), perception (by estimating the most likely state), and planning (by selecting the policy or action sequence that results in the lowest \emph{expected} free energy). 
        

        The expected free energy $G$ is a quantity that is only used during planning and represents the agent's variational free energy expected after pursuing a policy $\pi$. It is distinct from the variational free energy, because it requires considering future observations generated from the selected policy, as opposed to the current (and past) observations.        
        
        Active inference realizes goal-directed behavior by selecting policies that minimize expected free energy -- and that are expected to yield observations that are closer to preferred observations (or prior preferences). This is done by setting the approximate posterior of the policy proportional to this quantity: 

        \begin{equation}
            Q(\pi) = \sigma(\gamma G(\pi) + \log E),
        \end{equation}

        where $\sigma$ is the softmax function, $\gamma$ is a temperature variable, and $E$ is a prior distribution over actions, or habit. Actions are sampled according to this posterior distribution, where low temperatures yield more deterministic behavior. At a given timestep $\tau$, $G$ is computed for a given policy according to the following quantity as formalized in~\citep{heins_pymdp_2022}: 
        \begin{align}
            G(\pi, \tau) &= - \mathbb{E}_{Q(\mathbf{o}_\tau|\pi)}\big[D_{KL}[Q(\mathbf{s}_\tau|\mathbf{o}_\tau,\pi)||Q(\mathbf{s}_\tau|\pi)]\big] - \mathbb{E}_{Q(\mathbf{o}_\tau|\pi)}\big[\log P(\mathbf{o}) \big] \\
            &\quad+ \underbrace{\mathbb{E}_{Q(\mathbf{o}_\tau|\pi)}[D_{KL}(Q(\mathbf{s}_\tau|\mathbf{o}_\tau)||P(\mathbf{s}_\tau|\mathbf{o}_\tau,\pi))]}_{\text{Expected Approximation Error} \geq 0} \label{eq:approxerror}  \\
            &\geq - \underbrace{\mathbb{E}_{Q(\mathbf{o}_\tau|\pi)}\big[D_{KL}[Q(\mathbf{s}_\tau|\mathbf{o}_\tau,\pi)||Q(\mathbf{s}_\tau|\pi)]\big]}_{\text{Epistemic value}} - \underbrace{\mathbb{E}_{Q(\mathbf{o}_\tau|\pi)}\big[\log P(\mathbf{o}) \big]}_{\text{Pragmatic Value}}
            \label{eq:efe}
        \end{align}
        This quantity is decomposed into an \emph{epistemic value} and a \emph{pragmatic value} \citep{friston_active_2017}. 
        When evaluating $G$, we compute the upper bound shown in Equation~\eqref{eq:efe}. We thus make the assumption that the expected approximation error (Equation~\eqref{eq:approxerror}) is zero~\citep{heins_pymdp_2022}. 
        The epistemic value computes the expected information gain (Bayesian surprise) between the prior $Q(\mathbf{s}_\tau|\pi)$ and posterior $Q(\mathbf{s}_\tau|\mathbf{o}_\tau,\pi)$. Intuitively, this quantity represents how much the agent expects the belief over the state to shift when pursuing this policy. The pragmatic value then measures the expected log probability of observing the preferred observation under the selected policy, intuitively computing how likely it is that this policy will drive the agent to its prior preferences. The full expected free energy for a finite time horizon $T$ is computed as $\sum_{\tau=1}^T G(\pi,\tau)$.
        
        \subsubsection{Hierarchical active inference}

        Here, we discuss how the active inference scheme introduced above is extended to realize hierarchical perception and planning, through bottom-up and top-down message passing between the two levels of the hierarchical generative model shown in Figure~\ref{fig:generativemodel}. 
        

        During perception, first, the low-level state $\mathbf{s}^1$ is inferred given observations $\mathbf{o}_t$ and level 1 actions $\mathbf{a}^1_t$ using the inference mechanism implemented in PyMDP~\citep{heins_pymdp_2022}. When the free energy of the lowest level reaches a pre-specified threshold, a message containing the most likely level 1 state $\mathbf{s}^1_t$ the agent is in is passed to the level above (level 2). This threshold is chosen to be reached when the level 1-preference is reached, and uncertainty is below a fixed level. When level 2 gets this bottom-up message, it queries the environment for the presence of reward and observes $\mathbf{r}_\tau$. This conjunction of observations, together with high-level action $\mathbf{a}^2$ is used to infer the belief over the level 2 state $Q(\mathbf{s}^2)$, in the same way as done for level 1.

        During planning, i.e. generating a sequence of actions that (should) drive the agent toward its goal, policies are generated in a top-down manner, in the sense that goals set at level 2 determine the prior preference (and then in turn the policy) at level 1. First, the level 2 policy is selected by sampling from the approximate posterior over $\pi^2$ proportional to its expected free energy. The action selected at this level then conditions the policy at level 1 $\pi^1$, for which in turn the approximate posterior $Q(\pi^1)$ is computed. The low-level action $\mathbf{a}^1_t$ is then sampled according to this distribution,  similar to the hierarchical generative model described by Friston et al.~\citep{friston_deep_2017}.

        As depicted in Figure~\ref{fig:generativemodel}, the policy of the lower level sets a limited amount of steps. In particular, when the preference of the lower level is reached, a message is passed to the upper level, transitioning it into the next state. In this way, inferring the policy at the lower level only considers one action of the upper level at a time. 

        The posterior over the level 2 policy $Q(\pi^2)$ is first inferred, as it has no dependencies from above. To do this, the expected free energy is computed, for which we look one time step ahead in our implementation as this could predict the next corridor in which the agent encounters reward. We consider Equation~\ref{eq:efe2}, for which the observations are now a conjunction of reward $\mathbf{r}_\tau$ and level 1-states $\mathbf{s}^1$: 

        \begin{equation}
            G^2(\pi^2,\tau) \geq -\mathbb{E}_{Q(\mathbf{r}_\tau,\mathbf{s}^1_{t=\tau}|\pi^2)}\big[D_{KL}[Q(\mathbf{s}^{2}_\tau| \mathbf{r}_\tau,\mathbf{s}^1_t,\pi^2) ||Q(\mathbf{s}^2_\tau|\pi^2)]\big] -
            \mathbb{E}_{Q(\mathbf{r}_\tau,\mathbf{s}^1_{t=\tau}|\pi^2)}\big[\log P(\mathbf{r}_\tau,\mathbf{s}^1_{t=\tau}) \big],
        \label{eq:efe2}
        \end{equation}

        Note that the level 1 state is synchronized with the level 2 state, and only the state observed at this synchronized time is considered, hence we denoted the time index for the level 1 state by $t=\tau$. In practice, we used a temperature ($\gamma$) value of $0.5$ for this level. For the habit policy E, we use a categorical distribution over action, that is conditioned on the current state of the agent. The action $\mathbf{a}^2_\tau$ is then sampled according to $Q(\pi^2)$.

        The agent, however, is not able to act upon the physical world yet with this level 2 action. The agent now has to infer the posterior distribution over the policy at level 1 $Q(\pi^1)$. Because we now want to move towards a specific state, i.e. a disambiguated observation, we set the preference in state space for computing the expected free energy at level 1. In practice, Equation~\ref{eq:efe}, for level 1 then boils down to: 
        \begin{equation}
            G^1(\pi^1,t) \geq - \mathbb{E}_{Q(\mathbf{o}_t|\pi^1)}\big[D_{KL}[Q(\mathbf{s}^1_t|\mathbf{o}_t,\pi^1)||Q(\mathbf{s}^1_t|\pi^1)]\big]
            - \mathbb{E}_{Q(\mathbf{s}^1_t|\pi^1)}\big[\log P(\mathbf{s}^1_t) \big],
        \end{equation}

        where the first term yields the expected information gain after pursuing policy $\pi^1$, and the latter the expected utility. In other words, how close the expected state is from the preferred state, set by the action of the level above. This quantity is then used to approximate the posterior over the policy at level 1 from which action $\mathbf{a}^1_t$ is sampled. At this level, we set a temperature ($\gamma$) value of $0.5$, and a uniform habit distribution. Crucially, in order to evaluate $G^1$ at planning depth 1, we set the prior preference ($\mathbf{C}$ matrix) as such that the preference for each state is proportional to the distance to the goal, i.e. intuitively this means that the agent can follow a breadcrumb trail towards the goal, given that it properly inferred the current state. 
       
        \subsection{Casting CSCGs as partially observable Markov decision processes}
        \label{sect:map}

        CSCGs can be used directly to plan~\citep{george_clone-structured_2021}. However, in this work, we use the two learned CSCGs to create a hierarchical generative model for active inference -- and then use hierarchical active inference to solve the spatial alternation tasks. 

        Using the two CSCGs to create a hierarchical generative model for active inference requires mapping them into two partially observable Markov decision processes (POMDPs) \citep{kaelbling1998planning}, the mathematical framework used in discrete time active inference.

        \begin{figure}[t!]
            \centering
            \includegraphics[width=0.35\textwidth]{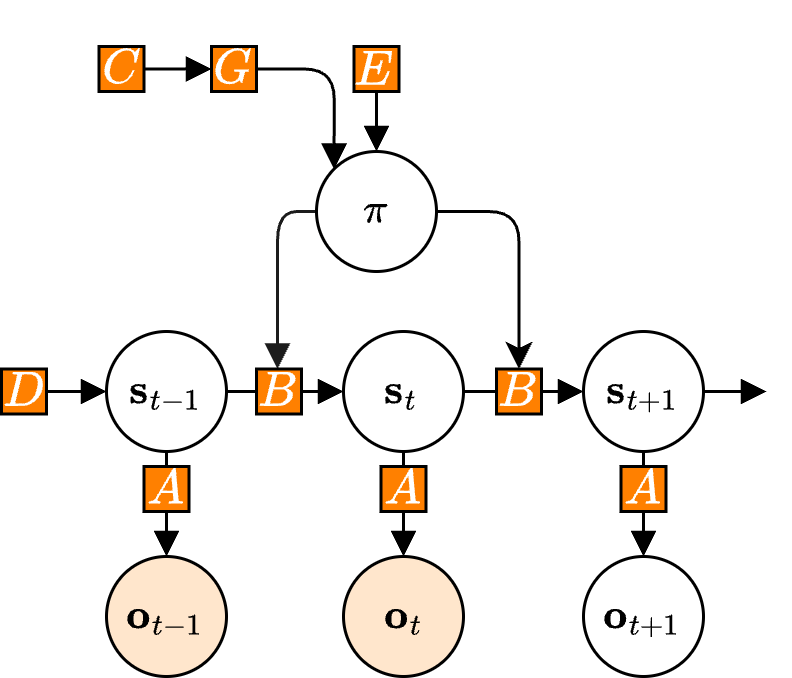}
            \caption{\textbf{Factor graph of a POMDP}: The conditional dependencies in Hidden Markov Models are parameterized by a set of matrices. The $\mathbf{A}$ matrix parameterizes the likelihood model, i.e. how states $\mathbf{s}_t$ map to observations $\mathbf{o}_t$. The $\mathbf{B}$ matrix parameterizes the transition model, i.e. how state $\mathbf{s}_{t+1}$ changes at each timestep, dependent on policy $\mathbf{\pi}$. The $\mathbf{C}$ vector denotes the preferred observation or state, and directly influences the $\mathbf{G}$ factor which conditions $\pi$. The policy $\pi$ depends on the habit $E$. Orange circles denote that the variable is observed.}
            \label{fig:factorgraph}
        \end{figure}

        
        Practically, a POMDP is described by a set of four arrays as shown in Figure~\ref{fig:factorgraph}. We describe these arrays using the symbol notation typically adopted in discrete time active inference \citep{parr_active_2022}. The $\mathbf{A}$ matrix encompasses the likelihood model $P(\mathbf{o}|\mathbf{s})$, or how states are mapped to observations. The $\mathbf{B}$ tensor entails the transition model $P(\mathbf{s}_{t+1}|\mathbf{s}_t,\mathbf{a}_t)$, or how states change over time, conditioned on the selected action. The $\mathbf{C}$ vector sets the prior over future observations or states, depending on the implementation. This vector is used within active inference to embed the preference or goal state/observation into the agent. Finally, the $\mathbf{D}$ vector describes the initial belief over the state $P(\mathbf{s})$. 


        As mentioned earlier, clone-structured cognitive graphs are a special case of hidden Markov models and can be therefore easily mapped to POMDPs (which extend HMMs, too). First, we consider the mapping of the likelihood. This $\mathbf{A}$ matrix represents the mapping of observation to state, specifically $\mathbf{A}_{i,j} = P(\mathbf{o}_i|\mathbf{s}_j)$. The matrix can be constructed through the definition of the CSCG, a deterministic mapping of clone state to its observation is set: $P(\mathbf{o}_i|\mathbf{s}_j)=1 \forall \mathbf{s}_j \in C(\mathbf{o}_i)$ and $P(\mathbf{o}_i|\mathbf{s}_j)=0 \forall \mathbf{s}_j \notin C(\mathbf{o}_i)$, where $C(\mathbf{o}_i)$ denotes all clone states for observation $\mathbf{o}_i$. When constructing the model we only consider states for which the marginalized probability $P(\mathbf{s}_j)$ surpasses a threshold of $0.0001$. Finally, we add an additional ``dummy'' state to the additional ``dummy'' observation, to which unlikely actions are mapped (see below). 

        To define the transition model, or $\mathbf{B}$ tensor, we use the learned parameters of the model. Through the learning process described in Section~\ref{sect:cscg}, the CSCG has learned the transition probability $P(\mathbf{s}_{t+1}|\mathbf{s}_{t},\mathbf{a}_t)$. This is parameterized as a tensor for which $\mathbf{B}_{i,j,k} = P(\mathbf{s}_i|\mathbf{s}_j,\mathbf{a}_k)$. We construct this tensor with the learned probabilities, for the same states that are considered in the likelihood matrix. Note that for ease of implementation, the active inference routines that we used require that any action could be executed from any state. However, in our task, some actions are not available or highly unlikely in some states (e.g., turning actions in a corridor). To handle this, we created a ``dummy state'' in the POMDP, which corresponds to observation 20 in Figure~\ref{fig:observations} to which we map the ``highly unlikely'' actions and to which we assign a large negative value (see below) to ensure that it is never selected during planning. In practice, we set a transition probability of $10^{-12}$ from every state to the dummy state (after which we re-normalize the tensor to sum to one for each action and state). We also set a self-transition of $1$ for the dummy state, therefore ensuring that it is ``absorbing''.
        


        We endow the active inference agent with prior preferences to secure rewards in the spatial alternation tasks in the $\mathbf{C}$ matrix. For ease of implementation, we set these prior preferences manually, rather than extracting them from the learned parameters of the CSCG. For level 2, we set a preference over the conjunction of observations where the reward is $1$ to encourage the agent to follow the learned rule. As described above, when the level 2 action is selected, it sets the $\mathbf{C}$ matrix of level 1 to a large preference for the preferred state and to a value proportional to the distance (in the \emph{learned} state space of level 1) from said given state to the preferred state (except for the dummy state, for which the $\mathbf{C}$ vector has a fixed value. This is set to the lowest value, making the dummy state the least preferred state). 

        The prior over the state is parameterized by the $\mathbf{D}$ matrix. We parameterize this as a uniform distribution over all the states, except for the dummy state. This reflects the fact that in our simulations, the active inference agent is placed in the W-maze with a random pose and has to self-localize.

        Finally, the habit, or the prior over action is parameterized by the $\mathbf{E}$ tensor. For level 1, we parameterize this as a uniform distribution. For level 2, we model this habit as a Dirichlet distribution, that is conditioned on the current state of the agent, and is proportional to actions that were rewarding during learning. 

        Using these tensors, inference at each single level can be implemented as a Bayes filter, iteratively computing the posterior over state at each timestep using the following update:
        \begin{align}
            \mathbf{qs}_t = \sigma(\mathbf{A} \cdot \mathbf{o}_t + \mathbf{B}_{\pi_{t-1}} \cdot \mathbf{qs}_{t-1}),
        \end{align}
        where $\mathbf{qs}_t$ denotes the parameters of the categorical distribution over state, i.e. $Q(\mathbf{s}_\tau|\mathbf{o}_\tau,\pi)=\text{Cat}(\mathbf{qs}^1_\tau)$, $\mathbf{o}_t$ represents the observation as a one-hot vector, $\sigma$ denotes the softmax function, and the $\mathbf{B}$-tensor is sliced by the policy $\pi_{t-1}$~\citep{friston_active_2017}. In the first timestep, $\mathbf{qs}_0$ is initialized as the prior matrix $\mathbf{D}$.
        
        The expected free energy $G$ for a considered policy $\pi_t$ can then be evaluated using these tensors for each future timestep $\tau$. As we specify the preference over state for the first level, and over observation for the second level, the distinction is made explicitly. For the first level, this boils down to~\citep{parr_active_2022}:   
        \begin{align}
            G^1_\tau &= \text{diag}(\mathbf{A}^{1}{}^\top \cdot \ln \mathbf{A}^1) \cdot \mathbf{qs}^1_\tau - \mathbf{qo}^1_\tau \cdot \ln \mathbf{qo}^1_\tau + \mathbf{qs}^1_\tau \cdot \ln \mathbf{C}^1, 
        \end{align}
        where the superscript $^1$ denotes the first level, $\mathbf{qs}^1_\tau$ denotes the parameters of the categorical distribution over state, i.e. $Q(\mathbf{s}^1_\tau|\mathbf{o}^1_\tau,\pi^1_t)=\text{Cat}(\mathbf{qs}^1_\tau)$, which is computed as $\mathbf{qs}^1_\tau = \mathbf{B}_{1,\pi^2_t} \cdot \mathbf{qs}^1_t$. The parameters of the categorical over the observation is denoted by $\mathbf{qo}^1_\tau$, i.e. $Q(\mathbf{o}^1_\tau)=\text{Cat}(\mathbf{qo}^1_\tau)$, and is computed as $\mathbf{qo}^1_\tau = \mathbf{A}^1 \cdot \mathbf{qs}^1_\tau$. 
        The same equation can be used for the expected free energy at the second level, except now the preference is specified as an expected observation: 
        \begin{align}
            G^2_\tau &= \text{diag}(\mathbf{A}^2{}^\top \cdot \ln \mathbf{A}^{2}) \cdot \mathbf{qs}^{2}_\tau - \mathbf{qo}^{2}_\tau \cdot \ln \mathbf{qo}^{2}_\tau + \mathbf{qo}^{2}_\tau \cdot \ln \mathbf{C}^{2}, 
        \end{align}
        where the superscript $^2$ denotes the second level. $Q(\mathbf{s}^2_\tau|\mathbf{o}^2_\tau,\pi^1_t)=\text{Cat}(\mathbf{qs}^2_\tau)$, which is computed as $\mathbf{qs}^2_\tau = \mathbf{B}_{2,\pi^2_t} \cdot \mathbf{qs}^2_t$. The parameters of the categorical over the observation at level 2 is denoted by $\mathbf{qo}^2_\tau$, i.e. $Q(\mathbf{o}^2_\tau)=\text{Cat}(\mathbf{qo}^2_\tau)$, and is computed as $\mathbf{qo}^2_\tau = \mathbf{A}^2 \cdot \mathbf{qs}^2_\tau$. 
        The full expected free energy $G$ can then be acquired by summing over time horizon $T$: $G = \sum_{\tau=t+1}^T G_\tau$, and the posterior over the policies is achieved through $Q(\pi) = \sigma(\gamma \cdot G + \ln E)$.

\section*{Acknowledgements}

This research received funding from the European Union’s Horizon 2020 Framework Programme for Research and Innovation under the Specific Specific Grant Agreements No. 945539 (Human Brain Project SGA3) and No. 952215 (TAILOR); the European Research Council under the Grant Agreement No. 820213 (ThinkAhead), the Italian National Recovery and Resilience Plan (NRRP), M4C2, funded by the European Union – NextGenerationEU (Project IR0000011, CUP B51E22000150006, “EBRAINS-Italy”; Project PE0000013, CUP B53C22003630006, "FAIR”; Project PE0000006, CUP J33C22002970002 “MNESYS”), and the PRIN PNRR P20224FESY. The GEFORCE Quadro RTX6000 and Titan GPU cards used for this research were donated by the NVIDIA Corporation. The funders had no role in study design, data collection and analysis, decision to publish, or preparation of the manuscript.


\bibliographystyle{apalike}
\bibliography{references}

\newpage
\appendix

\section*{Supplementary materials: Bridging Cognitive Maps: a Hierarchical Active Inference Model of Spatial Alternation Tasks and the Hippocampal-Prefrontal Circuit}

\section{Learning the spatial level using a Hidden Markov Model}

We investigate the benefit of using the CSCG's for learning the distinct layers of the hierarchy over using standard Hidden Markov Models on a navigation task. Specifically, we learn the spatial level using both approaches on the same dataset: a trajectory of the agent that contains a path from each distinct pose (position and orientation) to another. We consider the following variants, where the model is used by an active inference agent implemented in pymdp~\citep{heins_pymdp_2022}: (i) a CSCG agent, as used in the main text, (ii) a HMM agent trained using the Baum-Welch algorithm (on the same data), and (iii) a HMM fine tuned by the Viterbi algorithm (on the same data). 

The performance of each agent is measured as the success rate for reaching each corridor end from each considered starting pose in the maze within 30 steps. Since the observations are ambiguous, and the goals are specified in state space, the corresponding state is first extracted using the following mechanism: an agent is started in an unambiguous position (the T-junction) and the path to a corridor end is replayed, while the states are inferred. The resulting state is the state for which the agent has encoded this corridor end. From the results shown in Table~\ref{tab:spatialnavigation}, it is clear that only the CSCG is able to properly navigate the ambiguous W-maze. 

\begin{figure}
    \centering
    \begin{subfigure}{0.45\textwidth}
        \includegraphics[width=\textwidth]{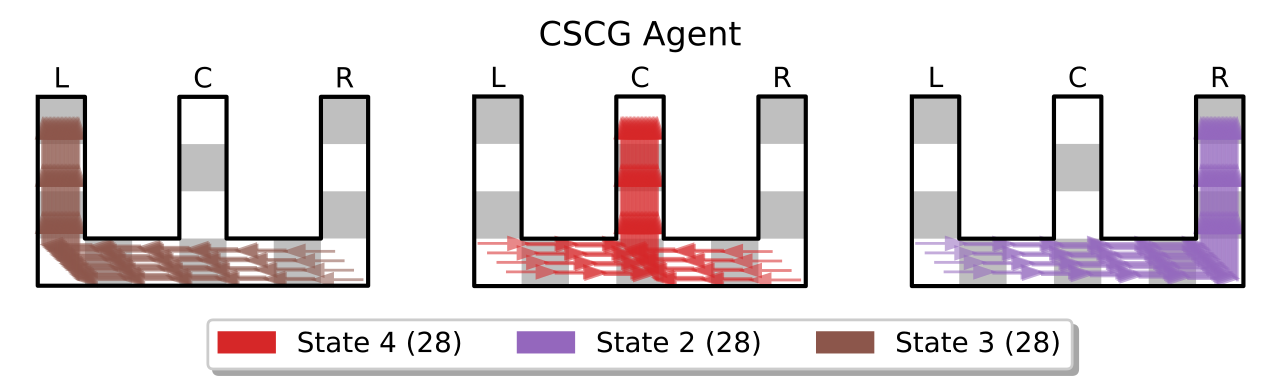}
        \subcaption{}
    \end{subfigure}
    \hfill
    \begin{subfigure}{0.45\textwidth}
        \includegraphics[width=\textwidth]{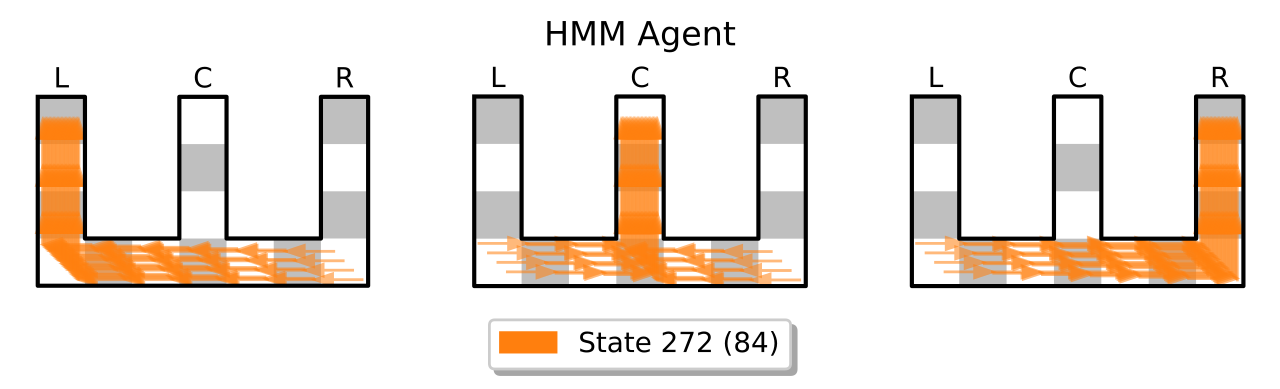}
        \subcaption{}
    \end{subfigure}
    \begin{subfigure}{0.45\textwidth}
        \includegraphics[width=\textwidth]{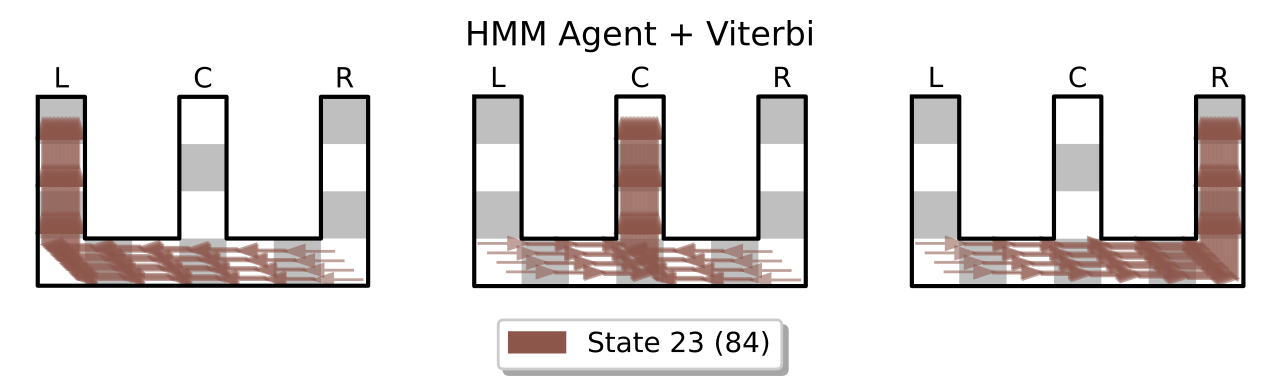}
        \subcaption{}
    \end{subfigure}
    \caption{\textbf{State disambiguation for the W-maze}: This figure displays the replayed trajectory from unambiguous locations in the bottom row of the maze toward each of the corridor ends. The trajectories are colored by the final inferred state, i.e. the state that encodes the end of the corridor.} 
    \label{fig:disambiguation}
\end{figure}

We determine the cause for the low performance of the HMM agents to be the inability to disambiguate the distinct corridor ends. As shown in Figure~\ref{fig:disambiguation}, this fails even when starting from unambiguous positions, e.g. the bottom row of the maze has some unique identifiers such as the T-junction and the 2 corners. In contrast, the CSCG encodes each corridor end in a distinct state.

\begin{table}[h!]
    \centering
    \caption{\textbf{Success ratio for spatial navigation:} The performance for reaching the three corridor ends as a goal location using the different models (CSCG, HMM, HMM+Viterbi) starting from each starting pose: 16 positions $\times$ 4 directions $\times$ 3 goals = 192 trajectories.}
    \label{tab:spatialnavigation}
    \begin{tabular}{c|c|c}
        & \textbf{success ratio} & \textbf{\% success} \\
        \hline
         CSCG Agent & 192/192 & 100.0 \% \\
         HMM Agent & 23/192 &  11.9\% \\
         HMM + Viterbi Agent & 32/192 & 16.7\% \\
    \end{tabular}
\end{table}

\section{Behavioral patterns observed in the W-maze} 

\begin{figure}[h!]
    \centering
    \begin{subfigure}{0.75\textwidth}
        \includegraphics[width=\textwidth]{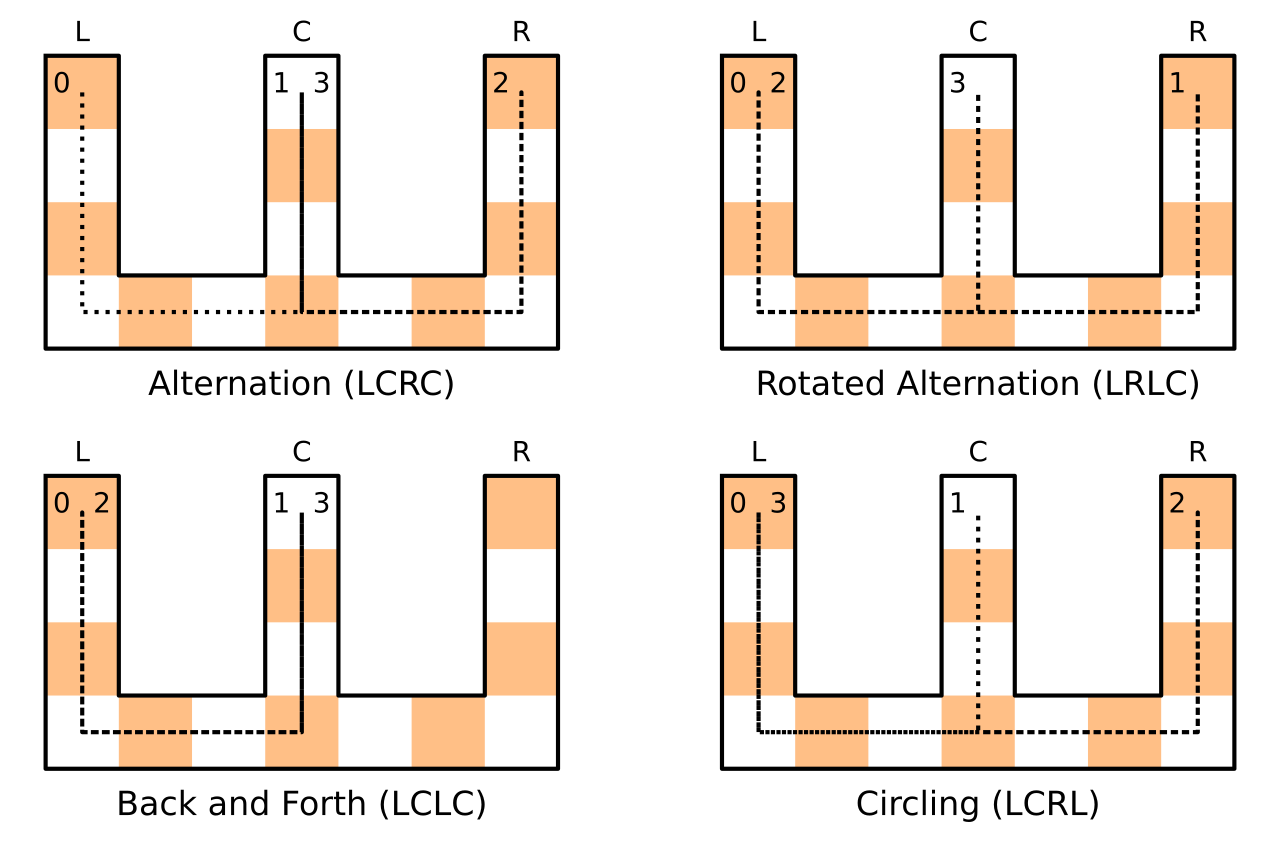}
        \subcaption{}
    \end{subfigure}
    \caption{\textbf{Behavioral patterns for the W-maze}: Visualization of multiple behavioral patterns measured over 4 visits of corridors. The dotted line indicates the trajectory, and the numbers in the corridor ends indicate the order of the visit. (i) alternation: following the rule, (ii) rotated alternation: following another rule that is currently not in play, (iii) back-and-forth: iterating between two corridors, and (iv) circling: cyclic evaluation of the different corridors. }
    \label{fig:error_patterns}
\end{figure}

The different behavioral patterns from~\cite{den_bakker_sharp-wave_2022} are depicted in Figure~\ref{fig:error_patterns}, specifically for the W-maze considered in this paper. There are four distinct patterns: (i) Alternation: the pattern of following a rule thet is in place. (ii) Rotated alternation: the pattern of following a rule that is currently not in place. (iii) Back and forth: iterating over two corridors, and (iv) Circling: visiting each corridor in a cyclic pattern. In both~\cite{den_bakker_sharp-wave_2022} and our study, these are measured over sub-sequences of four consecutive corridor visits. 

\section{Model specification}

The hyperparameters of the model are depicted in Table~\ref{tab:hyperparameters}. The threshold to determine when a message should be passed to the level above is $F_\text{threshold}^1$, and is related to the max-value $C_\text{max}$ used in the construction of the preference or $\mathbf{C}$ matrix. The policy length parameters determine how many steps are considered when evaluating the expected free energy.

\begin{table}[h!]
    \centering
    \caption{The hyperparameters of the hierarchical active inference model. The superscript indicates the level for which this parameter is used (e.g. $\gamma^1$ is the temperature for level 1.)}
    \label{tab:hyperparameters}
    \begin{tabular}{l|l}
        \textbf{Parameter} & \textbf{Value}\\ \hline
         $\gamma^1$ & 0.5  \\
         $\gamma^2$ & 0.5 \\
         $F_\text{threshold}^1$ & -14.8 \\
         $C_\text{max}^1$ & 15 \\
         $C_\text{max}^2$ & 1  \\
         $\text{policy length}^1$ & 5 \\
         $\text{policy length}^2$ & 2 \\
         $\text{n clones}^1$ & 20 \\
         $\text{n clones}^2 \text{(1 rule)}$ & 10  \\ 
         $\text{n clones}^2 \text{(3 rules)}$ & 32 
    \end{tabular}
\end{table}

\section{Control analyses for robustness}

\subsection{Sampling temperature $\gamma$}
\label{app:gamma}

\begin{figure}
    \centering
    \begin{subfigure}{0.35\textwidth}
        \centering
        \includegraphics[width=\textwidth]{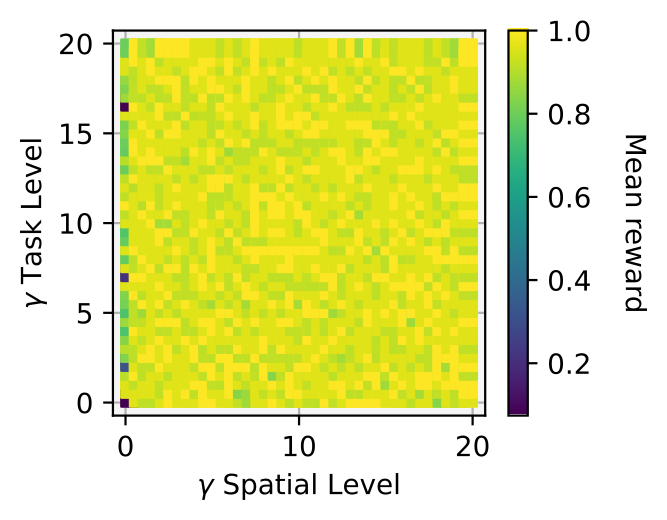} 
        \subcaption[]{\label{fig:sensitivity}}
    \end{subfigure}
    \begin{subfigure}{0.55\textwidth}
        \centering
        \includegraphics[width=0.95\textwidth]{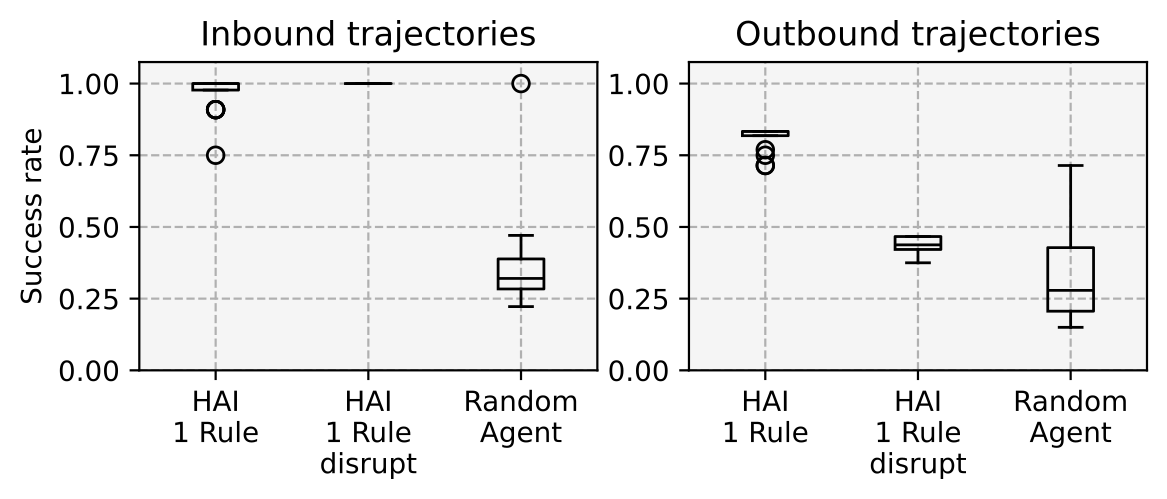}
        \subcaption[]{\label{fig:sim1_gamma16}}
    \end{subfigure}
    \caption{\textbf{Sensitivity analysis for $\gamma$}. (a) The mean reward acquired for different values of $\gamma$ in both the spatial and task level. We considered values from 0.5 to 20 in steps of 0.5. (b) Success rate for a hierarchical active inference (HAI), a disrupted HAI agent, and a random agent trained on a single rule. The values are computed over 20 trials, of 300 steps.}
\end{figure}

We performed a sensitivity analysis on the sampling temperature ($\gamma$ parameter) for both levels, as shown in Figure~\ref{fig:sensitivity}. We observed that the specific value of $\gamma$ does not have a large influence on performance, as assessed by mean reward collected.

We also conducted the same simulation as in Section~\ref{sect:sim1}, but with a sampling temperature of $16$. Figure~\ref{fig:sim1_gamma16} reports the mean success rate of visiting consecutive corridors while following the rule for an agent with and without disruption. This figure indicates that, without disruption, the agent is able to follow the rule and is near-perfect in collecting reward for both in- and outbound trajectories. When the disruption is applied, (i.e. the transition model of the highest level is impaired), the agent can still solve the inbound scenario, but drops to a chance-level performance for the outbound scenario. The higher value of $\gamma$ ensures that the agent leverages it's generative model more (i.e. the distribution over the next action has lower entropy) and is closer to greedily selecting the lowest expected free energy. This ensures that the agent will not select the central corridor to visit, as the generative model can predict that this will not yield any reward.

We note that the choice of temperature $\gamma$ has a large impact on the generated behavior (Table~\ref{tab:disruptionpatternsgam16}). When using a large value of $\gamma$, the agent resorts to greedy behavior (with respect to the expected free energy), while lower values allow the agent to sample according to the distribution over actions. We observe that with a temperature of 16, most of the generated behavior can be categorized in the back and forth category (83.77\%).

    \begin{table}[]
        \centering
        \caption{\textbf{Behavioral patterns during disruption:} Classified observed behavioral patterns according to~\cite{den_bakker_sharp-wave_2022} in different scenarios for agents trained on the three rules indicated by (3) and a single rule indicated by (1). Agent's policies are sampled using a sampling temperature $\gamma$ of 16.}
        \label{tab:disruptionpatternsgam16}
        \begin{tabular}{l|c|c|c|c|c}
            & \textbf{HAI} & \textbf{HAI} & \textbf{HAI} & \textbf{HAI }  & \textbf{Random}\\
            & \textbf{control (1)} & \textbf{disrupt (1)} & \textbf{control (3)} & \textbf{disrupt (3)} & \\
            \hline
            Alternation        &80.48\% & 61.80\% &  87.06\% & 0.00 \% &  3.47\%   \\
            Rotated alternation&0.00\%  & 0.00 \%  & 0.00\% & 0.00 \% &  8.24\%  \\
            Back and forth     &2.41\%  & 26.26\% &  4.31\% & 83.77\% &  8.86\%   \\
            Circling           &0.00\%  & 0.80 \%  & 4.85\% & 0.00 \% &  5.86\%  \\
            Other              &17.11\%  &11.14\%  & 3.77\% & 16.23\% &  73.75\% \\
        \end{tabular}
    \end{table}

\subsection{Control analysis: model capacity}
\label{sec:model_capacity}

We performed a robustness analysis of the model capacity (amount of parameters, length of the rules, and the amount of rules) by evaluating the average collected reward. The experiment considers 5 trials of 900 steps per trial, where the rule switches every 150 steps. The models are trained using the same hyperparameters from the main text, i.e. the amount of clones from Table~\ref{tab:hyperparameters}, and a sequence of 8000 corridor visits. In the case of multiple rules as in Experiment 2 (LCLR, LCRC, and RCRL), the rule switches every 1000 visits.

\begin{figure}
    \centering
    \begin{subfigure}{0.75\textwidth}
        \includegraphics[width=\textwidth]{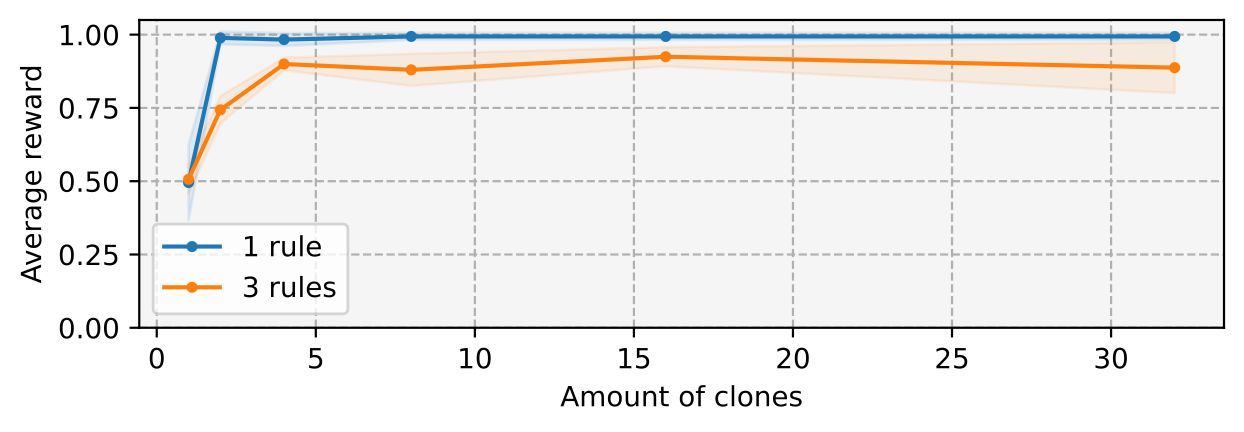}
        \subcaption[]{\label{fig:capacitynclones}}
        \includegraphics[width=\textwidth]{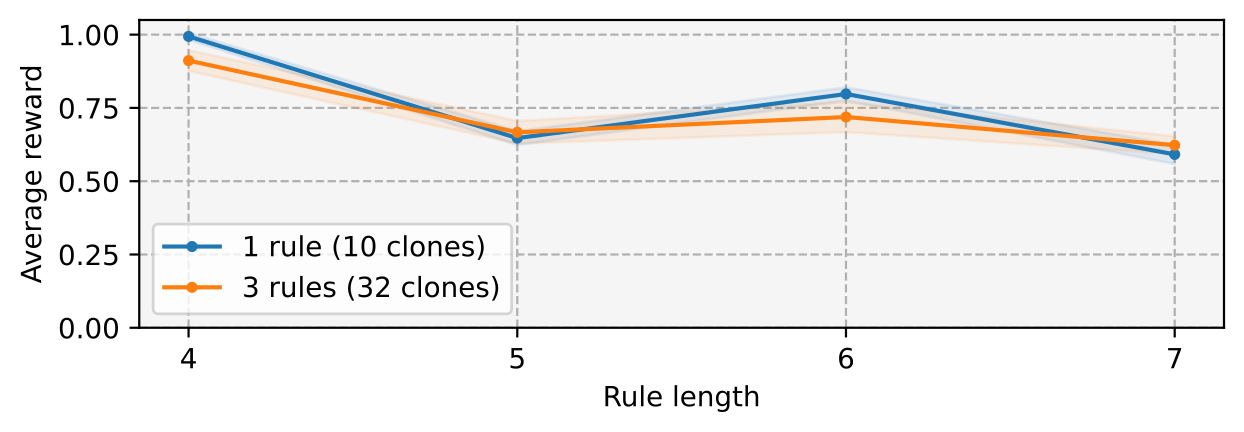}
        \subcaption[]{}
        \includegraphics[width=\textwidth]{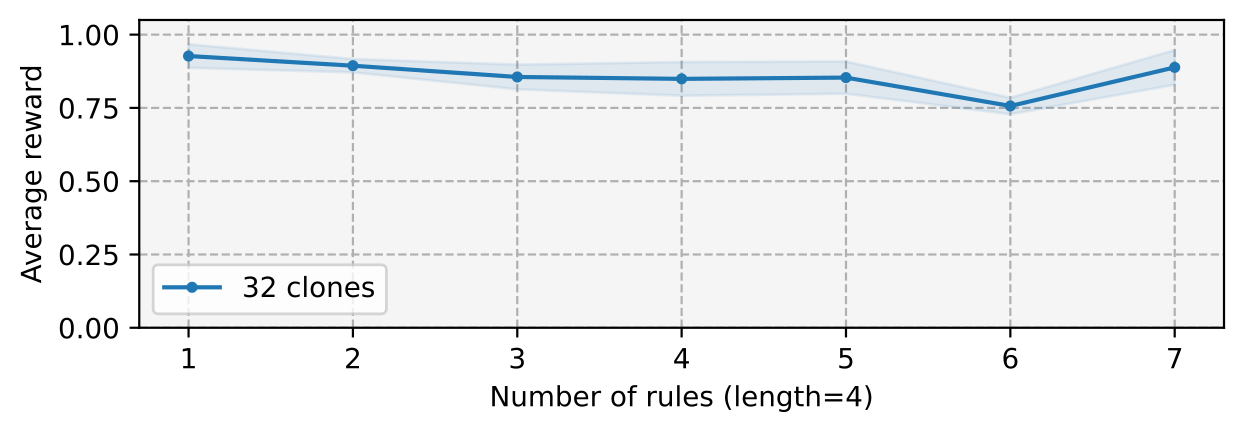}
        \subcaption[]{}
    \end{subfigure}
    \caption{\textbf{Capacity analysis of the hierarchical CSCG}. Each point is computed over 5 trials of length 900 steps, where the rule switches every 150 steps if multiple rules are in place. (a) The average collected reward in function of the number of clones for both the one and three rule case. (b) The average collected reward in function of rule length. (c) The average collected reward in function of the amount of learned rules. }
    \label{fig:capacityfig}
\end{figure}

In Figure~\ref{fig:capacitynclones}, the average collected reward is visualized in function of the amount of clone states, after a warmup period of 50 steps to infer the rule. We observe that the performance remains constant when we vary the amount of clones. We observe that for a single clone, the model does not have the capacity to learn the rule for both models. We observe that the model is able to learn (an average collected reward of over 80\%) the single rule starting at 4 states, and three rules from 8 clone states.

Next, we evaluated the impact of rule length. In the main text, all rules had a fixed length of four. We now lengthen this rule by adding more corridors (e.g. for the single rule case this becomes: 4: LCRC, 5: LCRCR, 6: LCRCRC, and 7: LCRCRCL). We observe a drop in performance when the rules become longer, indicating that the model does not have enough capacity to learn the rules. Note that in rule 5, the (R)ight corridor can be followed by a (L)eft or (C)enter corridor, while in the other rules this is always (C)enter, which could cause this change.

Finally, we evaluated the performance of our agent when multiple rules of length four are considered. For this experiment, we now consider a sequence of 1050 steps, where the rule switches every 150 steps. We consider an agent with 32 clones and vary the amount of learned rules between one and eight. We observe that the average collected reward remains similar.

\subsection{Control analysis: noisy scenario}
\label{sec:noisy_scelario}

We conducted a robustness study that evaluates how robust the active inference model is in case of a noisy environment. Consider the same W-maze where a single rule is in place, however, now whenever the agent enters a corridor, there is a 1/3 chance that the agent's prior beliefs (of both the navigation and prefrontal model) are reset to a uniform distribution. This means that, after this reset, the agent needs to infer what rule it is currently in, and where it currently is.

\begin{figure}[h!]
    \centering
    \hfill
    \begin{subfigure}{0.4\textwidth}
        \includegraphics[width=\textwidth]{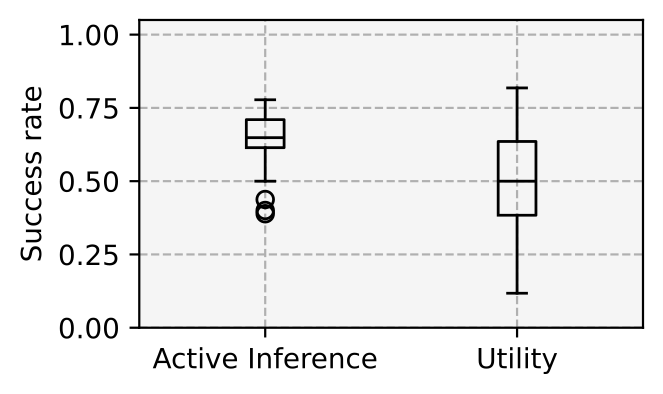}
        \subcaption[]{\label{fig:robust_success_rate}}
    \end{subfigure}
    \hfill
    \begin{subfigure}{0.4\textwidth}
        \includegraphics[width=\textwidth]{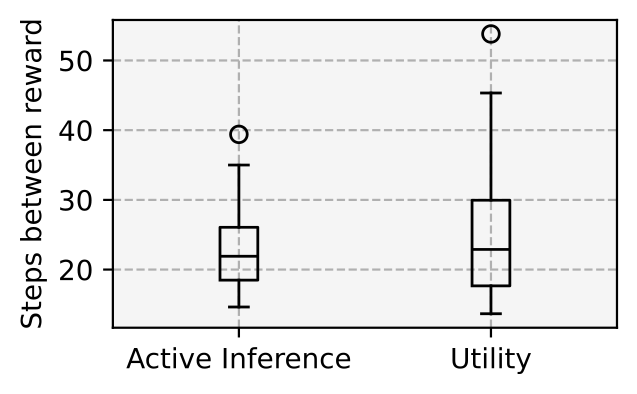}
        \subcaption[]{\label{fig:steps}}
    \end{subfigure}
    \hfill
    \caption{\textbf{Control Experiment: Robustness of the agent in a noisy scenario where the agent state is reset with a 33.3\% probability upon entering a corridor.} (a) The success rate of an active inference and a purely utility agent in visiting the corridors in the W-maze according to a fixed rule, in the noisy environment. The active inference agent has higher accuracy in the choice of the correct corridor according to the rule (t-test p=0.0147). The values are computed over 20 trials of 300 steps. (b) The average amount of steps between visiting a rewarding corridor. There is no significant difference between the visit times for both agents (t-test p=0.348) The values are computed over 20 trials of 300 steps. }
\end{figure}

We compared two agents: one that uses the expected free energy functional described in the main text (Active Inference agent) and another that only uses the utility and thus acts as a greedy agent (Utility agent). Figure~\ref{fig:robust_success_rate} shows that the average success rate (i.e. the ratio of corridor visits that yield reward) of the two agents trained on the same rule, over 300 trials, is significantly greater for the Active Inference agent compared to the Utility agent (p-value 0.0147). This is because while the Utility agent only seeks reward by reaching the end of corridors, the Active Inference agent seeks unambiguous locations to self-localize. Figure \ref{fig:steps} shows that the Active Inference agent also needs fewer steps to reach rewards, with a mean of 23.48 steps for the Active Inference agent and 26.3 steps for the Utility agent, but this difference does not reach significance. This is probably because this count also includes the steps required to reach unambiguous locations, which are greater for the Active Inference than for the Utility agent.

\end{document}